\documentclass{article}

 \usepackage{amssymb,amsmath}

\begin{document}

\def\S{{\mathbb S}}
 \def\R{{\mathbb R}}
 \def\C{{\mathbb C}}
\def\O{{\mathbf O}}
\def\K{{\mathbb K}}
\def\Z{{\mathbb Z}}
\def\F{{\mathbb F}}

\def\A{{\mathbb A}}

\def\Q{{\mathbb Q}}

\def\L{{ \Lambda}}

\def\cV{{\cal V}}
\def\cH{{\cal H}}
\def\cA{{\cal A}}
\def\cP{{\cal P}}
\def\cL{{\mathcal L}}
\def\cB{{\mathcal B}}
\def\cR{{\mathcal R}}
\def\cQ{{\mathcal Q}}

\def\frr{{\mathfrak r}}
\def\frI{{\mathfrak I}}
\def\frN{{\mathfrak N}}
\def\frF{{\mathfrak F}}
\def\frA{{\mathfrak A}}
\def\frK{{\mathfrak K}}
\def\frB{{\mathfrak B}}
\def\frb{{\mathfrak b}}
\def\frk{{\mathfrak k}}
\def\frL{{\mathfrak L}}
\def\frl{{\mathfrak l}}
\def\fri{{\mathfrak i}}
\def\frj{{\mathfrak j}}
\def\frm{{\mathfrak m}}

\def\mm{{\mathbf m}}
\def\nn{{\mathbf n}}
\def\kk{{\mathbf k}}
\def\ll{{\mathbf l}}

\def\phi{\varphi}
\def\kappa{\varkappa}
\def\epsilon{\varepsilon}
\def\le{\leqslant}
\def\ge{\geqslant}

\def\GL{{\rm GL}}
\def\Sp{{\mathbf {Sp}}}
\def\cl{\mathrm {cl}}

\def\wt{\widetilde}
\def\ov{\overline}

\def\rrr{\rightrightarrows}

\newcommand{\tr}{\mathop{\mathrm {tr}}\nolimits}
\newcommand{\Ker}{\mathop{\mathrm {Ker}}\nolimits}
\newcommand{\Indef}{\mathop{\mathrm {Indef}}\nolimits}
\renewcommand{\Im}{\mathop{\mathrm {Im}}\nolimits}

\newcommand{\Pol}{\mathop{\mathrm {Pol}}\nolimits}
\newcommand{\codim}{\mathop{\mathrm {codim}}\nolimits}

\newcounter{sec}
\renewcommand{\theequation}{\arabic{sec}.\arabic{equation}}
\newcounter{punkt}
\newcommand{\punkt}{\arabic{sec}.\arabic{punkt}. \addtocounter{punkt}{1}}
\newcounter{utver}
\newcommand{\utver}{\arabic{sec}.\arabic{utver}. \addtocounter{utver}{1}}

\begin{center}

{\bf \Large Structures of boson and fermion
  Fock spaces  in the space of symmetric
functions}

\medskip

{\sc Yurii A. Neretin}\footnote%
{Supported by grant NWO-047-008-009}

\end{center}

{\small We realize the Weil representation of infinite dimensional
symplectic group and spinor representation of infinite-dimensional
group $\GL$ by linear operators in the space of symmetric functions
in infinite number of variables.}

\bigskip

{\bf\punkt Purposes of this paper.}
Canonical unitary operators
connecting  a boson Fock space and a fermion Fock space
with a space of symmetric functions  are well known,
see \cite{PS},  \cite{MJD}
 and further references in these books.

The basic structure in the boson Fock
space is a semigroup of Gauss operators.
This semigroup contains the Friedrichs--Shale
 group of automorphisms
of the canonical commutation relations;
 see \cite{Ber1}, \cite{Ner-categories}).

The basic structure in the fermion Fock space is
a semigroup of Berezin operators
(i.e., fermion analogs of  Gauss operators).
This semigroup contains the Friedrichs--Berezin
 group of automorphisms
of the canonical anticommutation relations,
 see \cite{Ber1}, \cite{Ner-categories}).

 The purpose
of our paper is to transfer these structures
to the space of symmetric functions.
This problem is completely solved for the Gauss
operators (see below Subsection 3.5)
and partially solved for the Berezin
operators (Subsection 5.3).
Results of this paper were
 announced in \cite{Ner-symmetric}).

There are many problems of this kind, some of them
are discussed in \cite{Ner-poisson}, \cite{Ner-symmetric},
\cite{TsV}
.

\smallskip

{\bf\punkt Operators in the space of symmetric functions.}
We consider a Hilbert space $\S_\cl$, whose  elements
are formal symmetric series $f(x_1,x_2,x_3,\dots)$;
$\S_\cl$ is equipped with a classical
(Redfield) scalar product (see \cite{Mac}, I.4; see below 1.3).
For each formal series $K(x_1,x_2,\dots;y_1,y_2,\dots)$
symmetric separately in $x_j$ and in $y_j$,
we associate the linear operator $A$ in $\S_\cl$ given by
\begin{equation}
Af(x)=\langle K(x,y),\ov{f(y)}\rangle_{\S_\cl}
;
\end{equation}
here we consider $K(x,y)$ as a function in $y$
depending on the parameters $x$.
We say that $K(x,y)$ is the {\it kernel}
of the operator $A$.

\smallskip

{\bf\punkt the Weil representation.}
We realize  the 'Weil' representation  of the
Friedrichs--Shale symplectic group
(see its definition below 2.10)    by operators,
whose kernels have the form
\begin{multline}
\prod_{k<l}\Bigl\{1+ \sum_{i>0,j>0}a_{ij} x_k^i x_l^j\Bigr\}
\prod_{k,l}\Bigl\{1+ \sum_{i>0,j>0} b_{ij} x_k^i y_l^j\Bigr\}
\prod_{k<l} \Bigl\{1+ \sum_{i>0,j>0} c_{ij} y_k^i y_l^j\Bigr\}
\times\\ \times
\prod_k \Bigl\{1+\sum_{i>0}\alpha_i x_k^i\Bigr\}
\prod_k \Bigl\{1+\sum_{i>0}\beta_i y_k^i\Bigr\}
,
\end{multline}
where $a_{ij}=a_{ji}$, $c_{ij}=c_{ji}$.
Moreover, all the bounded operators in $\S_{cl}$, whose
kernels are given by expressions (0.2),
form a semigroup; this semigroup is isomorphic to the semigroup
of Gauss operators  in the boson Fock space.

The correspondence (0.1) between the kernels $K(x,y)$
and linear operators depends on a scalar product.
Many natural scalar products in the space of symmetric functions
are known (see \cite{Mac}, I.4, III.4, VI.5, V.10;
\cite{Ker}, \cite{GR}, \cite{Ner-hyper}, Section 10,
 \cite{Ner-symmetric}).  Our construction literally
survives for the Jack, Hall--Littlewood and Macdonald scalar
products; maximal generality, then it exists, is a family
of scalar product defined by Kerov in \cite{Ker}.

Nevertheless, in our context the classical case is distinguished,
since it is related to the Virasoro algebra (see \cite{PS});
for some formulae related to  actions of the group
of diffeomorphisms of the circle in $\S_\cl$, see
\cite{Ner-symmetric}.    On explicit description
of boson Fock spaces related to the Hall--Littlewood
and Macdonald  cases, see \cite{Ner-symmetric}.

 \smallskip

{\bf\punkt The spinor representation.}
The natural group $\mathbf O$ of symmetries
of the fermion Fock space is the group of
$(\infty+\infty)\times(\infty+\infty)$
 complex invertible  bounded
matrices $g=\begin{pmatrix}A&B\\C&D\end{pmatrix}$
such that $g$ is orthogonal
$$
\begin{pmatrix}A&B\\C&D\end{pmatrix}
\begin{pmatrix}0&1\\1&0\end{pmatrix}
\begin{pmatrix}A&B\\C&D\end{pmatrix}^t
=
                      \begin{pmatrix}0&1\\1&0\end{pmatrix}
$$
and $B$, $C$  are Hilbert--Schmidt matrices
(i.e., $\tr B^*B,\tr C^*C<\infty$);
  see \cite{Ner-categories}, IV.4,
this group is larger than the 'group of automorphisms of canonical
anticommutation relations' (see \cite{Ber1})
  described by Friedrichs, Bogolubov, Berezin
and Shale--Stinespring.

The representation of $\mathbf O$ in the fermion Fock space is
an infinite dimensional variant
of the spinor representation.

The space $\S_\cl$ is in a canonical one-to-one
correspondence with some subspace  in fermion Fock space
that is called '{\it space of semi-infinite  forms}',
see  definitions below in 4.2-4.3.
The natural group $\mathbf {GL}$ of symmetries of the space of semiinfinite
forms consists of
$(\infty+\infty)\times(\infty+\infty)$
 complex invertible  bounded
matrices
$\begin{pmatrix}P&Q\\R&T\end{pmatrix}$
such that
$Q$, $R$ are Hilbert--Schmidt matrices and
the Fredholm  index of $P$ is zero
(in our case, this is equivalent to the condition
 $\dim \Ker P=\codim \Im P$), see \cite{Ner-categories},
 IV.3--IV.4.

The group $\mathbf {GL}$ is a subgroup in   $\mathbf O$%
\footnote{In notation of \cite{Ner-categories},
our groups are the the following $(G,K)$-pairs:
$\mathbf {GL}=\bigl(
 \GL(2\infty,\C), \GL(\infty,\C)\times \GL(\infty,\C)\bigr)$
and $\mathbf O=\bigl(\mathrm O(2\infty,\C),\GL(\infty,\C)\bigr)$.
For description of the embedding
$\mathbf {GL}\to\mathbf O$, see \cite{Ner-categories}, IV.4);
to avoid misunderstanding, emphasis  that
      $\mathbf {GL}$ is a subgroup in the group
$\bigl(\mathrm O(4\infty,\C),\GL(2\infty,\C)\bigr)$
with the duplicated infinity.}.

Now consider a Laurent polynomial  in two    variables
$$
S(u,v)=\sum_{-M<i<\infty,\,-M<j<\infty} a_{ij} u^i v^j
$$
(only finite number of terms are nonzero).
Define the kernel $K(x,y)$
by the condition
\begin{multline*}
K(x,y)\Bigr|_{x_N=x_{N+1}=\dots=0,\,y_N=y_{N+1}=\dots=0}
=         \\ =
\frac{\det_{1\le k,l\le N}\{S(x_k,y_l)+\sum_{j=-M}^\infty x^j_k y^j_l\}
           \prod_{k=1}^N x_k^N y_k^N}
     {\prod_{1\le k<l\le N} (x_k-x_l)
      \prod_{1\le k<l\le N} (y_k-y_l)}
\end{multline*}
for all $N>M$.

We show that operators in $\S_\cl$
  with such kernels form an infinite-dimensional
group $\GL_\infty$,
 whose elements $g= \begin{pmatrix}P&Q\\R&T\end{pmatrix}$
satisfy the condition:
 $g-1$ has only finite number of nonzero matrix elements.

This is a partial answer to the question formulated above,
since this group $\GL_\infty$ is a proper subgroup in
the natural group of symmetries $\mathbf {GL}$.

\smallskip

 This construction exists in the space
$\S_\cl$  and  does not survive for general Kerov's scalar
products.

\smallskip

{\bf \punkt Structure of the paper.}
 Sections 1--2 contain preliminaries on
the space of symmetric functions and on the boson Fock
space. In Section 3, we discuss the boson-symmetric correspondences.

In section 4, we introduce a space of semiinfinite forms
and a space of skew-symmetric functions.
In Section 5, we discuss a fermion-symmetric correspondence.


\smallskip

{\bf Acknowledgements.}
This work was done during my visit to Yale University
in 1994. I  thank prof. R.Howe for hospitality
and discussions.
I also thank the organizers of Russian--French workshop
 on combinatorics  (Independent University of Moscow, May 2003),
 since this was an occasion for
writing the present paper.

\medskip

{\bf \large  1. Symmetric functions}

\nopagebreak

\medskip

\nopagebreak

\addtocounter{sec}{1}

\setcounter{equation}{0}
\setcounter{punkt}{1}
\setcounter{utver}{1}

{\bf \punkt Symmetric functions.}
In this paper, $x_1$, $x_2$, \dots is
an infinite collection of   formal variables.

We denote by $\ov\S=\ov\S(x)$ the space
of all formal series
in the variables $x_j$ symmetric with respect to permutations
of $x_j$.
We call  elements of $\ov\S$ by {\it symmetric functions}.

By $\S^k\subset\ov\S$ we denote the space of symmetric formal series
of degree $k$  in the variables $x_j$.
This space is finite dimensional, its dimension
equals  the number $p(k)$ of partitions of $k$.

By $\S\subset\ov\S$ we denote the space of series of bounded degree.

                  \smallskip

{\bf \punkt Some bases in $\S$.}  For details, see \cite{Mac},  I.2-3.

\nopagebreak

\smallskip

a) {\it Functions $p_\mm$.}
A Newton sum $p_m=p_m(x)$, where $m=1,2,\dots$, is
$$
p_m(x):=\sum_{j=1}^\infty x_j^m
.$$
Denote  by $\mm$ a collection
of nonnegative integers,
$$
\mm=(m_1,m_2,\dots);\qquad \text{$m_j=0$ for suffitiently large $j$.}
$$
Denote
$$p_\mm(x):=p_1(x)^{m_1} p_2(x)^{m_2} \dots$$
The functions $p_\mm$ form a basis in $\S$.

We also use  another notation for the same functions.
Let $\lambda$ be a sequence of integers
$$
\lambda:\,\,\lambda_1\ge\lambda_2\ge\dots,
\qquad \text{$\lambda_j=0$ for sufficiently large $j$.}
$$
Then
$$p_\lambda:=p_{\lambda_1}p_{\lambda_2}\dots.$$
In other words,
$p_\lambda(x)=p_\mm(x)$,
where $m_j$ is the number of entries of $j$ into
the collection  $\lambda$.

\smallskip

b) {\it Monomial symmetric functions  $m_\lambda$.}
Let
$\lambda=(\lambda_1,\lambda_2,\dots)$, where
$\lambda_1\ge\lambda_2\ge\dots$  are integers
and $\lambda_j=0$ for sufficiently large $j$.
Denote by $m_\lambda$
the sum of all distinct monomials of the form
$$
x_{k_1}^{\lambda_1}  x_{k_2}^{\lambda_2} x_{k_3}^{\lambda_3}
\dots
\qquad k_i\ne k_j.$$
Obviously, the system $m_\lambda$  is a basis in $\S$.

c) {\it Schur functions $s_\lambda=s_\frr$.}
Let $\lambda_1\ge\lambda_2\ge\dots$ be nonnegative integers,
and $\lambda_j=0$ for sufficiently large $j$.
The function $s_\lambda\in\S$ is defined by the rule
$$
s_\lambda(x_1,x_2,\dots)\Biggr|_{x_{n+1}=x_{n+2}=\dots=0}=
\frac{\det\limits_{1\le k\le n, 1\le j\le n}
   \bigl\{ x^{\lambda_j+n-j}_k\bigr\}}
    {\prod_{1\le k<l\le n}(x_k-x_l)}
\,.$$

We also use  another notation for the same functions.
Let $\frr$ be a sequence of integers such that
$$
\text{
$\frr:\,\,r_1>r_2>\dots$ and
$r_j=-j$ for sufficiently large $j$.}
$$
We assume
$$
s_\frr:=s_\lambda, \qquad \text{where $\lambda_j=r_j+j$}
.$$

{\sc Remark.} Thus, we introduced 3 types of notation for
Young diagrams; in these 3 cases we use respectively
Greek letters ($\lambda$, $\mu$, etc),
bold Latin letters ($\mm$, $\nn$, etc.),
and Gothic letters ($\frk$, $\frl$, etc.) as above.

\smallskip

{\bf \punkt Scalar products in $\S$.}
The {\it classical scalar product}
(J.H.Redfield, 1927, see \cite{Mac},  I.4)
in $\S$ is defined by the condition:
the functions $p_\mm$ are pairwise orthogonal and
$$
\|p_\mm\|^2=\prod m_j! j^{m_j}
.$$
The equivalent condition is:
the Schur functions $s_\lambda$
form an orthonormal basis.

\smallskip

More generally, let us define {\it Kerov's family}
$\langle\cdot,\cdot\rangle_\omega$ of scalar products
in $\S$. Fix a sequence
$$
\omega=(\omega_1, \omega_2,\dots);\qquad \text{$\omega_j>0$}
.$$
We assume that
the functions $p_\mm$ are pairwise orthogonal and
\begin{equation}
\|p_\mm\|^2=\prod m_j!\, \omega_j^{m_j}
.
\end{equation}

There are three following distinguished examples  of such
scalar products
\begin{align}
\omega_j&=j\alpha \qquad &\text{({\it Jack scalar products});}\nonumber\\
\omega_j&=j (1-t^j)^{-1}
       &\text{({\it Hall--Littlewood scalar products});}\nonumber\\
\omega_j&=j\cdot \frac{1-q^j}{1-t^j}
       &\text{({\it Macdonald scalar products}).}
\end{align}

We denote by $\S_\omega$ the completion of the Euclidean
space $\S$ equipped with the scalar product
$\langle\cdot,\cdot\rangle_\omega$.

More carefully, denote by $\S^\tau_\omega$
the (finite dimensional) Euclidean space $\S^\tau$
equipped with the Kerov scalar product.
Then $\S_\omega$ is a Hilbert direct sum
of the Hilbert  spaces  $\S^\tau_\omega$,
$$
\S^\tau_\omega=\oplus_{\tau=0}^\infty \S^\tau_\omega
.$$
Each element  $f\in\S_\omega$ can be represented as  a sum
of a series
$$
f=\sum_{\tau=0}^\infty f_\tau;\qquad
\text{where $f_\tau\in\S^\tau$,
and $\sum \|f_\tau\|^2_{\S^\tau_\omega} <\infty$}.
$$
In particular,
$\S_\omega\subset \ov\S$.
The scalar product is given by
$$
\langle  f, g \rangle_\omega =
\langle \sum f_\tau,\sum g_\tau \rangle_\omega :=
\sum
\langle f_\tau, g_\tau \rangle_{\S^\tau_\omega}\,.
$$

We emphasis that this  expression has sense also for
$f\in\ov \S$, $g\in \S$.

We also denote by $\S_\cl$ the space corresponding to the
classical case $\omega_j=j$.

\smallskip

{\bf \punkt Bisymmetric kernels.} Let $x_1$, $x_2$, \dots
and $y_1$, $y_2$, \dots  be two collections
of formal variables.
A {\it bisymmetric kernel} $K(x,y)$ is a formal series
symmetric with respect to $x_j$ and symmetric
with respect to $y_i$.

\smallskip

 {\bf \punkt Linear operators in $\S_\omega$.} Fix $\omega$.
Represent a bisymmetric kernel $K(x,y)$
as a series in $x_j$ with coefficients depending on $y_i$,
$$
K(x,y)=\sum x_1^{k_1} x_2^{k_2} x_3^{k_3}
 \dots u_{k_1,k_2,k_3,\dots} (y)
,$$
where $k_j$ are nonnegative integers, and all $k_j=0$
except a finite number of $j$. Then
$u_{k_1,k_2,k_3,\dots} (y)\in\ov\S $, these expressions also are symmetric
with respect to $k_j$. Hence
$$
K(x,y)=
\sum\limits_{\lambda_1\ge\lambda_2\ge\dots}
 m_\lambda(x)
  u_{\lambda_1,\lambda_2,\dots} (y)
.$$

We define the linear operator
$\cA_K $ in the space of symmetric functions by the formula
\begin{equation}
\cA_K f(x)=   \langle K(x,y), \ov {f(y)}\rangle_{\omega}:=
\sum\limits_{\lambda_1\ge\lambda_2\ge\dots}
 m_\lambda(x)
\langle
  u_{\lambda_1,\lambda_2,\dots}\, , \, \ov {f(y)}
\rangle_{\omega}
,\end{equation}
where $\ov f$ denotes the usual complex conjugation.

We also represent this expression in the form
\begin{equation}
\cA_K f=\sum_{\sigma=0}^\infty [\cA_K f]_\sigma
,\end{equation}
where
\begin{equation}
[\cA_K f]_\sigma :=
\sum_{\lambda_1\ge\lambda_2\ge\dots\, ,\,\, \sum\lambda_j=\sigma}
 m_\lambda(x)
\langle
  u_{\lambda_1,\lambda_2,\dots}\, , \, \ov {f(y)}
\rangle_{\omega}
.\end{equation}

Obviously, $[\cA_K f]_\sigma \in\S_\sigma$
and hence the summands of (1.4) are pairwise orthogonal.

\smallskip

{\sc  Proposition \utver}
{\it The map $K\mapsto \cA_K$ is a bijection of the space
of all bisymmetric kernels to the space of
all linear operators
$\S\to\ov \S$.}

\smallskip

{\sc Remark.} This bijection depends on $\omega$.

\smallskip

{\sc Proposition \utver} {\it Let $A:\S_\omega\to\S_\omega$
be a bounded operator. Then}

\smallskip

a) {\it There exists a bisymmetric kernel $K(x,y)$ such that
$Af=\cA_K f$ for $f\in\S$.}

\smallskip

b) {\it For $f\in \S_\omega$, its image $Af$ equals (1.4)--(1.5);
the series in the right hand side of (1.4)  converges
in the Hilbert space $\S_\omega$.}

\smallskip

{\sc Proposition \utver} {\it For each $\sigma=0,1,2,\dots$
choose an arbitrary  (in general, nonorthogonal)  basis
$\xi_1^{(\sigma)}$, \dots, $\xi_{p(\sigma)}^{(\sigma)}$
in $\S^\sigma$.
Represent a bisymmetric kernel $K(x,y)$ in the form
$$
K(x,y)=
\sum\limits_{\sigma=0}^\infty
\Bigl[\sum\limits_{i=1}^{p(\sigma)}
\xi_i^{(\sigma)}(x)   v^{(\sigma)}_i (y)
\Bigr]
,$$
where $v^{(\sigma)}_i $ are some elements of $\ov\S$.
  }

a) {\it For each $f\in\S$,
\begin{equation}
\cA_Kf(x)=
\sum\limits_\sigma\Bigl[ \sum\limits_i
\xi_i^{(\sigma)}(x)
\langle v^{(\sigma)}_i (y), \ov{f(y)}\rangle_\omega   \Bigr]
.\end{equation}

Summands of the formal series $\sum_\sigma$ do not depend on a choice
of a basis $\xi_i$.  In particular, they coincide with summands
of the series} (1.4).

b) {\it If $\cA_K$ is a bounded operator and $f\in \S_\omega$,
then $\cA_K f$ coincides with (1.6); the series
$\sum_\sigma$ in the right hand side
 converges in the Hilbert space $\S_\omega$.}

\smallskip

{\sc Proposition \utver}
{\it Let
$$
K(x,y)=\sum\limits_{\mm,\nn} \gamma_{\mm,\nn} p_\mm(x)\, p_{\nn}(y)
.$$
Then                  }
$$
\langle \cA_K p_\nn, p_\mm\rangle_\omega=
\gamma_{\mm,\nn} \prod_j \omega_j^{m_j+n_j} m_j!\, n_j!
.$$

\smallskip

{\sc Proposition \utver} {\it Let $\omega_j=j$, i.e., we have the
classical scalar product in $\S$.
Let
$$
K(x,y)=\sum\limits_{\lambda,\mu}
\beta_{\lambda,\mu}    s_\lambda(x) \,s_\mu(y)
.$$
Then}
$$
\langle \cA_K s_\mu, s_\lambda\rangle_\omega=
\beta_{\lambda,\mu}
.$$

{\sc Remark.} On the kernel of the identity operator, see
\cite{Mac}, I.4, see also below 3.6.

\smallskip

{\bf \punkt Proofs of Propositions 1.1--1.5.}
We start from Proposition 1.3a.
Represent $K(x,y)$ as the series
$$
K(x,y)=\sum_{\tau,\sigma}  K_{\sigma,\tau}  (x,y),
$$
where $K_{\sigma,\tau}$ has the degree $\tau$ in
the variables $x_1$, $x_2$, \dots
and the degree $\sigma$ in
the variables $y_1$, $y_2$, \dots.

Let $\cA_{\sigma,\tau}$ be the operator with the kernel
$K_{\sigma,\tau}(x,y)$,
$$
\cA_{\sigma,\tau}:\quad \oplus \S^j\to \oplus  \S^j
$$
Obviously, $\cA_{\sigma,\tau}$ is zero on all $\S^j$
for $j\ne\tau$,
and $\cA_{\sigma,\tau}:\,\S^\sigma\to \S^\tau$.

For the kernel $K_{\sigma,\tau}$ the statement
of Proposition 1.3a) is trivial. It is equivalent
to the following obvious lemma.

\smallskip

{\sc Lemma \utver} {\it Let $V$, $W$ be finite dimensional
Euclidean spaces.  Let $e_j$ be a basis
in $V$. Let $K\in V\otimes W$,
$$K=\sum e_j\otimes h_j,\qquad h_j\in W.$$
Then each linear operator $W\to V$ has the form
$$\cA_K w= \sum \ov{ \langle h_j, w\rangle_W}\, e_j $$
and $\cA_K$ does not depend on a choice  of the basis $e_j$.}

\smallskip

We apply this lemma to $W=\S^\tau_\omega$, $V=\S^\sigma_\omega$.

Thus, $\cA_K$ is a block operator
\begin{equation}
\cA_K=\begin{pmatrix} \cA_{11} & \cA_{12} &\dots   \\
                      \cA_{11} & \cA_{22} &\dots    \\
                      \vdots & \vdots & \ddots
      \end{pmatrix}
\end{equation}
in $\oplus \S^j$. Now Proposition 1.3a became obvious.

Let $\cA_K$ be bounded, let $f=\sum f_j$,
$Af=\sum g_j$, where $f_j$, $g_j\in\S^j$.
Then
$g_\tau=\sum \cA_{\tau,\sigma} f_\sigma$,
and this is the statement of Proposition 1.3b.

Thus Proposition 1.3 is proved.

\smallskip

If we assume $\xi=s_\lambda$, then we obtain Proposition 1.5.

If we assume $\xi=p_\mm$, then we obtain Proposition  1.4.
This implies Proposition 1.1.

If we assume $\xi=m_\lambda$ in Proposition 1.3.b,
then we obtain Proposition 1.2.

\smallskip

{\bf \punkt Products of operators.}

\nopagebreak

\smallskip

{\sc Proposition \utver}
{\it Let for bisymmetric kernels $K$, $L$ the operators
$\cA_K$, $\cA_L$ be bounded in $\S_\omega$. Then
$\cA_K \cA_L=\cA_M$, where}
\begin{equation}
M(x,z)=\langle K(x,y), \ov{ L(y,z)}\rangle_{\S_\omega(y)}
.\end{equation}

More exactly, we expand our kernels in $m_\lambda$
$$
K(x,y)=\sum_\lambda m_\lambda(x) u_\lambda(y);\qquad
L(x,y)=\sum_\mu m_\mu(z) v_\mu(y)
$$
and write
$$
M(x,z)=\sum_{\lambda,\mu} \langle u_\lambda ,\ov v_\mu\rangle_\omega
  m_\lambda(x) m_\mu(z)
$$

{\sc Proof.}
We represent $\cA_K$, $\cA_L$ as block matrices (1.7), and
the statement becomes obvious.
 We also can refer to Proposition 1.4.

\medskip

{\bf\large 2. Boson Fock space. Semigroup of Gauss operators}

\nopagebreak

\medskip

\nopagebreak

\addtocounter{sec}{1}

\setcounter{equation}{0}
\setcounter{punkt}{1}
\setcounter{utver}{1}

Here we discuss some basic definitions related to the boson
Fock space. For details, see \cite{Ner-categories}.

\smallskip

{\bf \punkt Boson Fock spaces with finite degrees of freedom.}
Let $n=0,1,2,\dots$. Consider the  space $\C^n$
with the coordinates $z_1$, $z_2$, \dots, $z_n$. Consider the space
$\Pol_n$ of polynomials in $z_j$; equip this space with
the scalar product
\begin{equation}
\langle f,g\rangle=\pi^{-n} \int_{\C^n} f(z)\ov {g(z)}
 \exp\{-|z|^2\}
\,\, |dz|
,\end{equation}
where $|dz|$ is the Lebesgue measure on $\C^n$.

The monomials $z_1^{m_1}\dots z_n^{m_n}$ are pairwise
orthogonal and
\begin{equation}
\|z_1^{m_1}\dots z_n^{m_n}\|^2=m_1!\dots m_n!
\end{equation}

The boson Fock space  $\F_n$ with $n$ degrees of freedom
is the completion of $\Pol_n$
with respect to the scalar product (2.1).
Elements of this space are entire functions
$f(z)$ on $\C^n$ satisfying
$$
\int_{\C^n}
|f(z)|^2       \exp\{-|z|^2\}\, |dz|    <\infty
.$$

{\bf \punkt The boson Fock space with infinite number
degrees of freedom.}
Let $f=f(z_1,\dots, z_n)\in \F_n$.
We define the function $I_nf\in \F_{n+1}$
as
$$
I_nf(z_1,\dots, z_n,z_{n+1})
=f(z_1,\dots, z_n)
$$
By (2.2), $I_n$ is an isometric embedding $\F_n\to \F_{n+1}$.

Consider the chain
$$
\F_0\subset \F_1\subset\dots \subset \F_n \subset\dots
$$
The boson Fock space  $\F=\F(z)$
with infinite  degree of freedom is  the completion of $\cup_n \F_n$.
The space $\F$ is called the {\it boson Fock space}.

Consider the system of functions
\begin{equation}
e_\mm(z):=\prod z_j^{m_j},
\end{equation}
where $m_j$ are nonnegative integers and $m_j=0$ for sufficiently
large $j$.
Obviously, these functions form an orthogonal basis in $\F$,
and
$$\|e_\mm\|^2=\prod m_j!.$$

Consider a series
$\sum c_\mm e_\mm(z)$  convergent   in the Hibert space $\F$.
It can easily be checked, that  this series
absolutely converges for each $z\in l_2$.
Hence we can consider elements of $\F$ as entire functions
on $l_2$.

Let $a=(a_1,a_2,\dots)\in l_2$.
Denote by $\phi_a(z)$ the function%
\footnote{There are many terms
that are used for this system of functions:
systems of coherent states, supercomplete basis,
overfilled basis, delta-functions.}
\begin{equation}
\phi_a(z) =\exp\{\sum z_j\ov a_j\}
.
\end{equation}
For each $f\in \F$,
 the following {\it reproducing property} holds
\begin{equation}
\langle f, \phi_a\rangle_\F= f(a)
.\end{equation}

{\bf \punkt Hilbert--Schmidt matrices.}
(see \cite{DS},XI.6,XI.9)
Recall that an infinite matrix $A=\{a_{ij}\}$
 is called a {\it Hilbert--Schmidt} matrix
if
$$
\tr A^*A=\sum_{ij} |a_{ij}|^2 <\infty
.$$

A matrix $A$ belongs to the {\it trace class} if
$$
\tr \sqrt{ A^*A} <\infty
.$$

Recall, that for Hilbert--Schmidt matrices
$A$, $B$, the matrix $AB$ belongs to the trace class.

Also, recall that for a  matrix $A$ of the trace class the determinant
$\det (1+A)$ is well-defined.

\smallskip

{\bf \punkt Gauss vectors.}
For a matrix $A$,
the symbol $\|A\|$ below denotes the norm of the operator $l_2\to l_2$,
i.e., $\|A\|^2$ is the maximal eigenvalue of $A^*A$.

Let $A=\{a_{ij}\}$ be a {\it symmetric}
(i.e., $a_{ij}=a_{ji}$)  Hilbert--Schmidt matrix and $\|A\|<1$.
Let $\alpha=(\alpha_1,\alpha_2,\dots)\in l_2$.
The corresponding {\it Gauss vector} $b[A|\alpha]$
is a function in the variable $z\in l_2$
 defined by
$$
b[A|\alpha](z)=
\exp\bigl\{ \frac 12 \sum a_{ij}z_i z_j +\sum \alpha_j z_j \bigr\}=
\exp\bigl\{\frac 12 z A z^t +\alpha z^t \bigr\}
.$$
Here ${}^t$ denotes the transposition, $z$, $\alpha$
 are considered as  matrix--rows.

{\sc Proposition \utver}
a) $b[A|\alpha]\in \F$.
$$
b) \qquad
\langle b[A|\alpha], b[B|\beta] \rangle_\F=
\det(1-A\ov B)^{-1/2}
\exp\Bigl\{\frac 12 (\alpha\,\, \ov \beta)
\begin{pmatrix} -A & 1\\ 1&-\ov B \end{pmatrix}^{-1}
\begin{pmatrix} \alpha^t\\ \ov \beta^t\end{pmatrix}
\Bigr\}
.$$

{\sc Remark.}
$$
\begin{pmatrix} -A & 1\\ 1&-\ov B \end{pmatrix}^{-1}
=
\begin{pmatrix}
   \ov B(1-A\ov B)^{-1}& (1-\ov B A)^{-1}\\
        (1-A\ov B)^{-1}& A(1-\ov B A)^{-1}
.\end{pmatrix}
$$

{\sc Remark.} Since $\|AB\|<1$, we define
$$
(1-A\ov B)^{-1/2}:=1+\frac12 A\ov B
+\frac 18 (A\ov B)^2+\dots
$$
and $\det(1-A\ov B)^{-1/2}:=\det[(1-A\ov B)^{-1/2}]$.

{\bf \punkt Operators.}
Let $H$ be a bounded operator $\F\to \F$.
Let
$$He_\nn=\sum_\mm h_{\mm,\nn} e_{\mm}.$$
Consider the function $K(z,u)$ ({\it kernel of the operator}) $H$
 on $l_2\times l_2$
given by
$$
 K(z,u)=
\sum_{\mm, \nn}
 h_{\mm,\nn}\nn! e_{\mm}(z)\,e_\nn(u)
,
$$
where $e_\nn$ is given by (2.3) and $\nn!=\prod n_j!$.
This series converges on $l_2\times l_2$, moreover
for a fixed $z\in l_2$, the function $k_z(u)=K(z,u)$
as a function in $u$ belongs to $\F$.

We also have
\begin{equation}
\langle  e_\mm, H e_\nn\rangle=h_{\mm,\nn} \mm!\nn!
\end{equation}

The operator $H$ can be reconstructed from $K(z,u)$
by the formula
$$
Hf(z)= \langle f(u), K(z,u)\rangle_{\F(u)}
:=\langle f(u), k_z(u)\rangle_{\F(u)}
.$$

If $K_1$ is the kernel of $H_1$ and $K_2$ is the kernel
of $H_2$, then the kernel of $H_1 H_2$ is
$$
L(z,w)=\langle K_1(z,u), K_2(u,w)\rangle_{\F(u)}
$$

{\bf \punkt Gauss operators.}
Let $A$, $B$, $C$ be infinite matrices. Let
$S:=\begin{pmatrix} A&B\\B^t&C\end{pmatrix}$.
Assume that these matrices satisfy the following conditions.

\smallskip

$0^\circ$.  $A$, $C$ are  symmetric matrices, i.e., $A=A^t$, $C=C^t$.
Equivalently,  $S$ is symmetric.

$1^\circ$.
$\|S\|\le 1$.

\smallskip

$2^\circ$.
          $\|A\|<1$, $\|C\|<1$;

\smallskip

$3^\circ$. The matrices  $A$, $C$  are  Hilbert--Schmidt.

\smallskip

Under these conditions,
we define the {\it Gauss operator}
$$
\cB\begin{bmatrix} A&B\\B^t&C\end{bmatrix}
:\F\to \F
.$$
 being the operator with the kernel
$$
K(z,u)=\exp\left\{\frac 12
\begin{pmatrix} z& u \end{pmatrix}
\begin{pmatrix} A&B\\B^t&C\end{pmatrix}
\begin{pmatrix} z^t\\ u^t \end{pmatrix}
\right\}
.$$

The conditions $1^\circ-3^\circ$ are necessary but not
sufficient for boundedness.
There are two simple sufficient conditions of boundedness.

\smallskip

{\sc Theorem \utver} (\cite{Ner-boson})
a) {\it If $A$, $C$ are  operators of the trace class, then
$\cB[\cdot]$
is bounded.}

\smallskip

b) {\it If
$
\left\|
\begin{pmatrix} A&B\\B^t&C\end{pmatrix}
\right\|< 1
$,
then
$\cB[\cdot]$
is bounded.}

A product of two Gauss operators is well-defined%
\footnote{for the case of unbounded operators, see
\cite{Ner-categories},VI.2}
 and it is given by
the following theorem.

\smallskip

{\sc Theorem \utver} (\cite{Ols}, \cite{NNO})
\begin{multline}
\cB\begin{bmatrix} A&B\\B^t&C\end{bmatrix}
\cB\begin{bmatrix} U&V\\V^t&W\end{bmatrix}=\\=
\det(1-CU)^{-1/2}
\cB\begin{bmatrix}
A+BU(1-CU)^{-1}B^t&B(1-UC)^{-1}V\\
V^t(1-CU)^{-1}B^t & W+V^t(1-CU)^{-1}CV^t\end{bmatrix}
.\end{multline}

{\bf \punkt Linear relations.}
Formula (2.7) hides a simple algebraic structure,
namely a product of linear relations.
Recall necessary definitions.

Let $V$, $W$ be linear spaces.
A {\it linear relation} $P:V\rrr W$
is a subspace $P\subset V\oplus W$.

Let $P:V\rrr W$, $Q:W\rrr Y$ be linear relations.
Their product $QP$ is a linear relation $V\rrr Y$
defined in the following way.
A vector $v\oplus y \in V\oplus Y$ is an element
of $QP$ if there exists $w\in W$ such that
$v\oplus w\in P$, $w\oplus y\in Q$.

\smallskip

{\sc Example.} Let $L: V\to W$ be a linear operator.
Its graph is a linear relation. A product of
operators corresponds to the product
of their graphs in the sense of linear relations.

\smallskip

For a linear relation $P: V\rrr W$,
we define its {\it kernel} $\Ker P\subset V$
and its {\it indefiniteness}
$\Indef P\subset W$ by
$$
\Ker P=P\cap V;\qquad \Indef P=P\cap W
.$$

\smallskip

{\bf \punkt Clarification of Theorem 2.3.}
For a given matrix
$S:=\begin{pmatrix} A&B\\B^t&C\end{pmatrix}$
we consider the subspace
$\cP[S]\subset [l_2\oplus l_2] \oplus [l_2\oplus l_2]$
consisting of vectors
$$
[v_+\oplus v_-] \oplus [w_+\oplus w_-]
$$
satisfying  the equations
$$
\begin{pmatrix} v^+\\ w^-\end{pmatrix}
=
\begin{pmatrix} A&B\\B^t&C\end{pmatrix}
\begin{pmatrix} v^-\\ w^+\end{pmatrix}
.$$
We consider this subspace as a linear relation
$l_2\oplus l_2 \rrr  l_2\oplus l_2$.

\smallskip

{\sc Theorem \utver} (\cite{NNO}, \cite{Ner-boson})
{\it The linear relations  $\cP[S]$, where $S$ satisfies conditions
$1^\circ$--$3^\circ$, form
a semigroup  with respect
to the multiplication of linear relations.
 The equality
$$\cB[S_1] \,\cB[S_2]=\mathrm{const}\cdot \cB[S_3]
.$$
is equivalent to}
$$
\cP[S_1]\,\cP[S_2]=\cP[S_3]
.$$

We call the semigroup of all the linear relations
$\cP[S]$ by {\it the symplectic semigroup}.

Now we intend to characterize linear relations
of type $\cP[S]$.

\smallskip

{\bf \punkt Geometric description
of symplectic semigroup.}
Denote by $V^+$, $V^-$, $W^+$, $W^-$
four copies of the space $l_2$. Let
$$V=V^+\oplus V^-,\qquad W=W^+\oplus W^-.$$
We define in $V$ the skew-symmetric bilinear
form
$\{\cdot,\cdot\}_V$ and the Hermitian
form
     $[\cdot,\cdot]_V$
by
\begin{align*}
\bigl\{\xi^+\oplus\xi^-,\eta^+\oplus\eta^-\bigr\}_V:=
\sum_j (\xi_j^+\eta_j^- - \xi_j^-\eta_j^+);\\
\bigl[\xi^+\oplus\xi^-,\eta^+\oplus\eta^-\bigr]_V:=
\sum_j (\xi_j^+\, \ov\eta_j^{\,+} - \xi_j^- \,\ov\eta_j^{\,-}).
\end{align*}
We define two forms in $W$ by the same formulae.
Further we define the skew symmetric  form
$\{\cdot,\cdot\}_{V\oplus W}$ and the Hermitian
form     $[\cdot,\cdot]_{V\oplus W}$
in ${V\oplus W}$ by
\begin{align*}
\bigl\{ v\oplus w, v'\oplus w'\bigr\}_{V\oplus W}
=
\bigl\{ v, v'\bigr\}_{V}-
\bigl\{ w,  w'\bigr\}_{ W}
.\\
\bigl[ v\oplus w, v'\oplus w'\bigr]_{V\oplus W}
=
\bigl[ v, v'\bigr]_{V}-
\bigl[ w,  w'\bigr]_{ W}.
\end{align*}

Our conditions for the matrix $S$ are equivalent to the following
conditions for linear relations $\cP[S]$.

\smallskip

 A) Our matrix $S=S^t$ is symmetric. This means that $\cP[S]$
is Lagrangian%
\footnote{A subspace $P$ is Lagrangian with respect
to a form $\{\cdot,\cdot\}$ if $\{v,w\}=0$ for each
$v$, $w\in P$ and $P$ is a maximal among subspaces having this
property.}
 with respect to the skew-symmetric form
$\{\cdot,\cdot\}_{V\oplus W}$.

 B) The condition $\|S\|\le 1$ is equivalent to the condition
$$
\bigl[v\oplus w, v\oplus w\bigr]_{V\oplus W}\ge 0
\qquad \text{for all $v\oplus w\in \cP[S]$}
.$$

 C) The conditions $\|A\|\le 1$, $\|C\| \le 1$
follow from $\|S\|\le 1$. The additional condition
$\|A\|< 1$   is equivalent to positive definiteness
of $[\cdot,\cdot]_V$ on $\Ker \cP[S]$.
The condition
$\|C\|< 1$   is equivalent to negative definiteness
of $[\cdot,\cdot]_W$ on $\Indef \cP[S]$.

 D) The condition $3^\circ$ from 2.6 is not so transparent
from geometrical point of view.
The orthogonal projectors $P\cap(V_-\oplus W)\to W_-$
and  $P\cap(V\oplus W_+)\to V_+$ must be Hilbert--Schmidt operators.

      \smallskip

The unit element of the symplectic semigroup
is the graph of the identity operator; the corresponding matrix
$S$ is $S=\begin{pmatrix}0&1\\1&0\end{pmatrix}$.
The group of automorphisms of the canonical commutation relations
defined in the next subsection is the group of invertible
elements of the symplectic semigroup.

\smallskip

{\bf \punkt The symplectic group.} Consider the
group $\mathrm{Sp}(2\infty,\R)$
 of bounded invertible  operators $g:l_2\oplus l_2 \to l_2\oplus l_2$
satisfying the following properties.

\smallskip

1) The operator  $g$ commutes with the $\R$-linear transformation
$v_1\oplus v_2\mapsto \ov v_2\oplus \ov v_1$,
where $v\mapsto \ov v$ is the coordinate-wise complex conjugation.
Equivalently, the matrix of $g$ has the following block form
\begin{equation}
g=\begin{pmatrix} P & Q\\
                 \ov Q & \ov P
  \end{pmatrix}
.
\end{equation}
Equivalently, $g$ preserves $\R$-linear subspace
$V_\R$ consisting of all the vectors having the form
$(v,\ov v)$.

2)
  The operator $g$ preserves the skew-symmetric form
$\{\cdot,\cdot\}$. Equivalently,
$$
g\begin{pmatrix} 0&1\\-1&0\end{pmatrix} g^t=
 \begin{pmatrix} 0&1\\-1&0\end{pmatrix}
.$$

3)
  The operator $g$ preserves the Hermitian form
$[\cdot,\cdot]$. Equivalently,
$$
g\begin{pmatrix} 1&0\\0&-1\end{pmatrix} g^*=
 \begin{pmatrix} 1&0\\0&-1\end{pmatrix}
.$$

{\sc Remark.} Any two of  conditions 1)--3)
imply the third condition.
The conditions 1) and 2) mean that $g$ preserves
a skew-symmetric bilinear form in the real linear space $V_\R$.

\smallskip

The Friedrichs--Shale (see \cite{Sha}, \cite{Ber1})
 group  $\Sp$ of {\it automorphisms
of canonical commutation relations}%
\footnote{It is the $(G,K)$-pair
$\bigl(\mathrm{Sp}(2\infty,{\mathrm U}(\infty)\bigr)$
in the notation of \cite{Ner-categories}.}
 consists of elements  $g\in\mathrm{Sp}(2\infty,\R)$,
such that the block $Q$ is a Hilbert--Schmidt matrix.

\smallskip

{\sc Proposition \utver} \cite{Ber1}
{\it For $g\in\Sp$, define the operator
$$
\rho(g):=\cB\begin{bmatrix}
       Q \ov P^{-1} & (P^*)^{-1}\\
       \ov P^{-1}& -\ov P^{-1} Q
   \end{bmatrix}
$$
Then $\rho(g)$ is a projective representation
of $\Sp$ and the operators $\det(PP^*)^{-1/4}\rho(g)$
are unitary.}

\medskip

{\bf\large 3.  Boson-symmetric correspondences}

\nopagebreak

\medskip

\nopagebreak

\addtocounter{sec}{1}

\setcounter{equation}{0}
\setcounter{punkt}{1}
\setcounter{utver}{1}

In this section, we consider arbitrary spaces $\S_\omega$.

{\bf \punkt Definition of the correspondence.}
For $f=f(z_1,z_2, \dots)\in \F$,
we define the element $\Theta f\in\S_\omega$  by
\begin{equation}
\Theta f(x_1,x_2,x_3,\dots)=
f(\omega_1^{-1/2} p_1(x),\,  \omega_2^{-1/2} p_2(x),  \,
 \omega_3^{-1/2} p_3(x), \dots)
.
\end{equation}

{\sc Theorem \utver} {\it The map
$\Theta$ is a unitary operator $\F\to\S_\omega$.}

\smallskip

{\sc Proof.} This follow from formulae (1.1), (2.2).
\hfill $\square$

\smallskip

{\sc Remark.} The square roots in formula (3.1) can be easily
deleted after minor variation of definition of boson Fock space.
This variant of a language is used in \cite{Ner-symmetric}.
Also, the boson Fock spaces corresponding to classical scalar products
(Redfield, Hall--Littlewood, Macdonald) are described in
\cite{Ner-symmetric}.

\smallskip

{\bf \punkt Inversion formula.}

\nopagebreak

\smallskip

{\sc Theorem \utver} {\it Let $g\in\S_\omega$.
Then
$$
\Theta^{-1} g(z)=
\langle g, \Phi_z\rangle_\omega
,$$
where $\Phi_z(x)\in\S_\omega$ is given by}
$$
\Phi_z(x)=\prod_k
\exp\bigl\{
  \sum\limits_{j=1}^\infty
     \ov z_j \omega_j^{-1/2} x_k^j
\bigr\}
.$$

{\sc Proof.} The functions $\Phi_a(x)\in\S_\omega$
 correspond to the elements  $\exp(\sum a_j z_j)\in \F$,
see (2.4).
It remains to apply the reproducing property (2.5).
More carefuly,
$$f(z):=\langle f,\phi_z\rangle_\F=
\langle \Theta f, \Theta \phi_z\rangle_\omega=
\langle \Theta f, \Phi_z\rangle_\omega
$$


\smallskip

{\bf \punkt Correspondence of operators.}

\nopagebreak

\smallskip

{\sc Proposition \utver} {\it Let $A:\F\to\F$
be a bounded operator, let $K(z,u)$ be its kernel.
Then the bisymmetric kernel of the operator
$$\Theta A \Theta^{-1}:\,\,\S_\omega\to\S_\omega$$
is given by                         }
$$
K(\omega_1^{-1/2} p_1(x),\, \omega_2^{-1/2} p_2(x),
\dots;\,
\omega_1^{-1/2} p_1(y),\, \omega_2^{-1/2} p_2(y),
\dots)
,$$

{\sc Proof.}  See Proposition 1.4 and formula (2.6).

\smallskip

{\bf \punkt Multiplicative vectors.}
Let $A=\{a_{ij}\}$ be a symmetric matrix
($i,j>0$),
let $\alpha=\{\alpha_j\}$ be a sequence.
We define the formal series
$\Psi[A | \alpha]\in\ov\S$ by
\begin{multline}
\Psi[A | \alpha](x)=
\exp\Bigl\{\frac 12
\sum\limits_{ 1\le i<\infty,\,1\le j<\infty  }
a_{ij}
 p_i(x) \,p_j(x) +\sum\limits_{1\le j<\infty} \alpha_j p_j(x)
\Bigr\}
=\\=
\prod_{1\le  k <l<\infty}
\exp\Bigl\{\sum _{ 1\le i<\infty,\,1\le j<\infty  }
  a_{ij} x_k^i x_l^j\Bigr\}\cdot
\prod_{1\le k<\infty}
\exp\Bigl\{\sum _{ 1\le j<\infty}
\bigl(\alpha_j+\sum_{\sigma+\tau=j} a_{\sigma\tau}\bigr)\, x^j_k
\Bigr\}
.\end{multline}

{\sc Remark.} Consider an arbitrary formal series
having  the form
\begin{equation}
\prod_{1\le  k <l<\infty} R(x_k,x_l)
\prod_{1\le k<\infty} Q(x_k)
,
\end{equation}
where
$Q(x)=1+\sum_{j> 0} q_j x^j$ and
$$
R(x,y)=    1+
\sum\limits_{i> 0,\, j> 0}
       r_{ij} x^i y^j;\qquad  r_{ij}=r_{ji}
$$
Each series (3.3) can be represented in the form (3.2).
\hfill $\square$

\smallskip

Denote
\begin{equation}
\Omega:=\begin{pmatrix}
\omega_1&0&\dots\\
0&\omega_2&\dots\\
\vdots& \vdots& \ddots
\end{pmatrix}
.\end{equation}

{\sc Theorem \utver} a) {\it Denote $\wt A:=\Omega^{1/2} A \Omega^{1/2}$.
Let the matrix $\wt A$ be  Hilbert--Schmidt, and $\| \wt A\|<1$.
Let $\alpha \Omega^{1/2}\in l_2$. Then
$\Psi[A|\alpha]\in\S_\omega$.}

\smallskip

b) {\it Let $A$, $\alpha$ and $B$, $\beta$ satisfy the conditions
of statement a). Then}
\begin{multline*}
\langle \Psi [A|\alpha], \Psi[B|\beta]\rangle_\omega
=\\=
\det(1-A\Omega \ov B\Omega)^{-1/2}
\det\Bigl\{
\begin{pmatrix} \alpha & \ov\beta \end{pmatrix}
\begin{pmatrix} -A & \Omega^{-1} \\
               \Omega^{-1} & -\ov B \end{pmatrix}^{-1}
  \begin{pmatrix} \alpha^t \\ \ov\beta^t \end{pmatrix}
  \Bigr\}
\end{multline*}

{\sc Proof.} The function $\Psi[A|\alpha]$ is the image of
the Gauss vector
$$b\bigl[\Omega^{1/2} A \Omega^{1/2}|\,\alpha\Omega^{1/2}\bigr]$$
under the map $\Theta$. Thus the statement follows
from  Proposition 2.1.
\hfill $\square$

\smallskip

{\bf \punkt Multiplicative bisymmetric kernels.}
We intend to discuss bisymmetric kernels of the type
\begin{equation}
\prod\limits_{k<l} P(x_k,x_l) \prod\limits_{k,l} Q(x_k,y_l)
\prod\limits_{k<l} R(y_k,y_l)
\prod\limits_k \pi(x_k) \prod\limits_k \rho(y_k)
,\end{equation}
where $P$, $Q$, $R$, $\pi$, $\rho$ are formal series.

Fix symmetric matrices $A=\{a_{ij}\}$, $C=\{c_{ij}\}$
 and a matrix $B=\{b_{ij}\}$, where $i,j>0$.
Assume that the matrix
$$
S:=
\begin{pmatrix} \Omega^{1/2} & 0\\
                0& \Omega^{1/2}
\end{pmatrix}
\begin{pmatrix} A& B\\ B^t &C\end{pmatrix}
\begin{pmatrix} \Omega^{1/2} & 0\\
                0& \Omega^{1/2}
\end{pmatrix}
$$
satisfies the conditions of Subsection 2.6, i.e.,

\smallskip

$1^*$. $\|S\|\le 1$.

\smallskip

$2^*$.  $\|\Omega^{1/2} A\Omega^{1/2}\|<1$,
 $\|\Omega^{1/2} C\Omega^{1/2}\|<1$.

\smallskip

$3^*$.  $\Omega^{1/2} A\Omega^{1/2}$, $\Omega^{1/2} C\Omega^{1/2}$
        are Hilbert--Schmidt matrices.

\smallskip

Then we define the bisymmetric kernel
$\frK\begin{bmatrix} A& B\\ B^t &C\end{bmatrix}  (x,y)$
by
\begin{multline}
\frK\begin{bmatrix} A& B\\ B^t &C\end{bmatrix} (x,y)
=\\=
\prod\limits_{k<l}
  \exp\Bigl\{\sum_{i>0,j>0} a_{ij} x^i_k x^j_l\Bigr\}
\prod\limits_{k,l}
  \exp\Bigl\{\sum_{i>0,j>0} b_{ij} x^i_k y^j_l\Bigr\}
\prod\limits_{k<l}
  \exp\Bigl\{\sum_{i>0,j>0} c_{ij} y^i_k y^j_l\Bigr\}
\times\\ \times
\prod_{k\ge 1}\exp\Bigl\{\sum_j
   \bigl(\sum\limits_{\sigma,\, \tau:\,\,\sigma+\tau=j}
      a_{\sigma\tau}\bigr)  x_k^j \Bigr\}
\prod_{l\ge 1}\exp\Bigl\{\sum_j
   \bigl(\sum\limits_{\sigma,\,\, \tau:\,\sigma+\tau=j}
      c_{\sigma\tau}\bigr)  y_l^j \Bigr\}
\end{multline}

The same kernel can be also represented in the form
\begin{equation}
\frK\begin{bmatrix} A& B\\ B^t &C\end{bmatrix} =
\exp\Bigl\{
\frac 12\sum_{i,j} a_{ij} p_i(x) \,p_j(x) +
\sum_{i,j} b_{ij} p_i(x) \,p_j(y) +
\frac 12\sum_{i,j} c_{ij} p_i(y) \,p_j(y)
\Bigr\}
\end{equation}

Denote by $\frB\begin{bmatrix} A& B\\ B^t &C\end{bmatrix} $
the operator $\S_\omega\to \S_\omega$
defined by the kernel
 $\frK\begin{bmatrix} A& B\\ B^t &C\end{bmatrix} $.

\smallskip

{\sc Theorem \utver} {\it Let the conditions $1^*$--$3^*$
are satisfied.}

\smallskip

a)  {\it Let $\|S\|<1$.
Then the operator
$\frA\begin{bmatrix} A& B\\ B^t &C\end{bmatrix}  $ is bounded.}

\smallskip

b) {\it Let $\Omega^{1/2} A\Omega^{1/2} $,
   $\Omega^{1/2} C\Omega^{1/2} $ be operators of the trace class.
       Then $\frA\begin{bmatrix} A& B\\ B^t &C\end{bmatrix}$ is bounded.}

\smallskip

{\sc Theorem \utver}
{\it Let $\frA\begin{bmatrix} A& B\\ B^t &C\end{bmatrix}$,
$\frA\begin{bmatrix} U& V\\ V^t &W\end{bmatrix}$
be bounded operators. Then
\begin{multline*}
\frA\begin{bmatrix} A& B\\ B^t &C\end{bmatrix}
\frA\begin{bmatrix} U& V\\ V^t &W\end{bmatrix}
=\det(1-C\Omega U\Omega)^{-1/2}
\times \\ \times
\frA\begin{bmatrix}
   A+B\Omega U (\Omega^{-1}-C\Omega U)^{-1} B^t&
   B(\Omega^{-1}-U\Omega C)^{-1} V \\
   V^t (\Omega^{-1}-C\Omega U)^{-1} B^t &
   W+V^t(\Omega^{-1}-C\Omega U)^{-1} C\Omega V
  .\end{bmatrix}
\end{multline*}
The semigroup of operators $\frA[\cdot]$ is isomorphic
to the semigroup of Gauss operators $\frB[\cdot]$.}

\smallskip

{\sc Proofs.} By Proposition 3.3,
$$\frA \begin{bmatrix} A& B\\ B^t &C\end{bmatrix}
=
\Theta \circ
\cB
\Bigl[
\begin{pmatrix} \Omega^{1/2} & 0\\
                0& \Omega^{1/2}\end{pmatrix}
\begin{pmatrix} A& B\\ B^t &C\end{pmatrix}
\begin{pmatrix} \Omega^{1/2} & 0\\
                0& \Omega^{1/2}\end{pmatrix}
\Bigr]
\circ
\Theta^{-1}
.$$
Now our theorems follow from Theorems 2.2, 2.3.

\smallskip

{\bf  \punkt Example. The identity operator.}
The identity operator in $\F$ has the kernel
$$\exp\bigl\{\sum z_j u_j\bigr\}.$$
Hence the bisymmetric kernel of the identity
operator in $\S_\omega$ is
$$
K(x,y)=
\prod_{k,l} \exp\Bigl\{\sum_{j>0} \omega_j^{-1} x_k^j y_l^j \Bigr\}
.$$

1) For the Redfield scalar product, i.e., $\omega_j=j$, we obtain
$$
K(x,y)=\prod\nolimits_{k,l} (1-x_k y_l)^{-1}
.$$

2) For the Jack scalar product, i.e., $\omega_j=j\alpha$,
we have
$$
K(x,y)=\prod\nolimits_{k,l} (1-x_k y_l)^{-1/\alpha}
.$$

3) For the Macdonald scalar  products (see (1.2)),
$$
\sum_{j>0} \frac {x^jy^j}{\omega_j}=
\sum_{j>0} \frac 1j \cdot \frac {1-t^j} {1-q^j} x^jy^j
.$$
Expanding
$$
\frac {1-t^j} {1-q^j}= \sum_{k\ge 0} q^{jk} - t^j\sum_{k\ge 0} q^{jk}
,$$
we obtain
$$
\sum_{j>0} \frac {x^jy^j}{\omega_j}=
-\sum_{k\ge 0}\ln(1-q^kxy)+  \sum_{k\ge 0}\ln(1-q^k t xy)
.$$
Finally,
$$
K(x,y)=
\prod_{k\ge 0} \frac {1- txyq^k}   {1- xyq^k}
.$$

All these formulae are well known, see \cite{Mac},
Sections I.4, III.4, VI.2, VI.10, see also \cite{Ker}.

\smallskip

{\bf \punkt Example. The Heisenberg group.}
Formula (3.5) is more general than (3.6). Our construction
can be easily extended to this more general case,
it is sufficient to apply   \cite{Ner-categories}, VI.4;
the semigroup of all bounded operators in $\S_\omega$ with kernels (3.5)
is isomorphic to the semigroup of all bounded operators in $\F$
having kernels of the form
$$
\exp\Bigl\{\frac 12\begin{pmatrix} z&u\end{pmatrix}
\begin{pmatrix} K&L\\L^t&M\end{pmatrix}
\begin{pmatrix} z^t\\ u^t\end{pmatrix} +
z\alpha^t+u\beta^t\Bigr\}
$$

We do not discuss the general case
and consider an example.

Let $\alpha$, $\beta\in l_2$.
Consider the operator $T(\alpha, \beta)$ in $\F$
whose the kernel  is
$$
\exp\bigl\{\sum z_j u_j + \sum z_j \alpha_j + \sum u_j\beta_j\bigr\}
.$$
Then (see \cite{Ner-categories}, Section VI.4)
$$
T(\alpha, \beta) T(\alpha', \beta')=\exp\bigl\{\sum \alpha'_j\beta_j\}
T(\alpha+\alpha',\beta+ \beta')
.$$
Thus, the operators $T(\alpha, \beta)$
form the complex Heisenberg group.

\smallskip

{\sc Remark.} If $\beta_j=-\ov\alpha_j$, then the operators
$T(\alpha, \beta)$ are unitary up to a scalar factor.
Otherwise, they are unbounded. Nevertheless
our operators
and their products are well-defined;
 for details, see \cite{Ner-categories}, VI.4.
\hfill $\square$

\smallskip

Let us describe the corresponding construction in $\S_\omega$.
Let  $a\Omega^{1/2}$,  $b\Omega^{1/2}\in l_2$.
Consider the operator $R(a,b)$ in $\S_\omega$,
whose bisymmetric kernel is
$$
\prod_{k,l}\exp\Bigl\{\sum_{j>0} \omega_j^{-1}x^j_k y^j_l\Bigr\}
\prod_k \exp\Bigl\{\sum_{j>0} a_j x^j_k\Bigr\}
\prod_l \exp\Bigl\{\sum_{j>0} b_j y^j_l\Bigr\}
.$$
Then
$$
R(a,b) R(a',b')= \exp\bigl\{\sum \omega_j b_j a'_j\bigr\}
R(a+a',b+b')
.$$
These operators form the complex Heisenberg group.

\smallskip

{\sc Remark.} The Heisenberg algebra (creation-annihilation operators)
consists of operators with kernels
$$
\prod_{k,l}\exp\Bigl\{\sum_{j>0} \omega_j^{-1}x^j_k y^j_l\Bigr\}
\cdot\bigl(r+ \sum_k\sum_j s_j x^j_k+ \sum_k\sum_j t_j y^j_k\bigr)
$$

\smallskip

{\bf\punkt Example: a formula for kernels of  operators.}
We intend to write a formula reconstructing
the bissymmetric kernel $K(x,y)$ by the
operator $A$.

Let $x_j$, $y_j$, $u_j$ be 3 collections of formal variables.
Then
\begin{equation}
K(x,y)=\langle
\prod_{k,l} \exp\Bigl\{\sum_{j>0} \omega_j^{-1} x_k^j u_l^j \Bigr\}
, A_u
\prod_{k,l} \exp\Bigl\{\sum_{j>0} \omega_j^{-1} u_k^j y_l^j \Bigr\}
\rangle_{\S_\omega[u]}
\end{equation}
In this formula, $A_u$ means that the operator $A$ acts on
symmetric functions in the variables $u_j$
 depending on the parameters  $y_j$. Then we consider
scalar product of functions in the variables $u_j$, depending
on the parameters $x_j$, $y_j$; see explanations to formula (1.8) above.

To prove (3.8), we expand
$$
\prod_{k,l} \exp\Bigl\{\sum_{j>0} \omega_j^{-1} u_k^j y_l^j \Bigr\}
=\sum_\nn \frac{\omega^\nn}{\nn!}p_\nn(u)p_\nn(y)
$$
and apply Proposition 1.4.

\medskip

{\bf\large 4. Space of semiinfinite forms $\L$
 and space of skew-symmetric functions}

\nopagebreak

\medskip

\nopagebreak

\addtocounter{sec}{1}

\setcounter{equation}{0}
\setcounter{punkt}{1}
\setcounter{utver}{1}

By $S_\infty$ (respectively $S_{2\infty}$)
  we denote the group of all finite
permutations of the set $\{1,2,3,\dots\}$
(respectively
$\{-2,-1,0,1,2,3,\dots\}$).

\smallskip

{\bf\punkt Anticommuting variables.} Let $\xi_1$, \dots, $\xi_n$
be anticommuting variables, i.e.,
\begin{equation}
\xi_i \xi_j =- \xi_j \xi_i
\end{equation}
for all pairs $i$, $j$; in particular $\xi_j^2=0$.
Denote by $\Lambda_n$ the space of all polynomials
in $\xi_j$; evidently,
$\dim\Lambda_n=2^n$.
We assume that the monomials
$\xi_{i_1}\xi_{i_2}\dots \xi_{i_k} $,
where $i_1>i_2>\dots>i_k$, form an orthonormal basis
in $\Lambda_n$.

Consider a family of linear forms
$a_1(\xi)$, \dots, $a_m(\xi)$ and
$b_1(\xi)$, \dots, $b_m(\xi)$:
$$a_k(\xi)=\sum_{j=1}^n  a_{kj}\xi_j;\qquad
  b_k(\xi)=\sum_{j=1}^n  b_{kj}\xi_j
.$$
Denote by $A$ and $B$ the matrices $\{a_{ij}\}$, $\{b_{ij}\}$.

\smallskip

{\sc Lemma \utver}    {\it Let $m\le n$. Then}
\begin{align*}
a)&\qquad a_1(\xi)\dots a_m(\xi)=\sum_{j_1>j_2> \dots> j_m}
    \det_{1\le i,s,\le m} \{a_{ij_s}\} \xi_{j_1}\dots \xi_{j_m}
\\
b)&\qquad
\langle a_1(\xi)\dots,a_m(\xi), b_1(\xi)\dots b_m(\xi)  \rangle
=\det\{\langle a_k,b_l\rangle\}=
\det AB^*
\\
c)& \qquad
\prod_{l=1}^m\bigl(\sum_{k=1}^m h_{lk} a_k(\xi)\bigr)
=\det\{h_{kl}\} \prod_{l=1}^m a_l(\xi)
\end{align*}

These statements are obvious.

\smallskip

{\bf \punkt The space $\L$.}  (see \cite{Ber2}, \cite{FF})
Consider a countable collection
$$\dots, \xi_{-2}, \xi_{-1}, \xi_0, \xi_1, \xi_2,\dots$$
of anticommuting variables,

A {\it semi-infinite monomial} is an infinite product of the type
$$
\Xi_\frk=\prod_{i=1}^\infty \xi_{k_j}:=
\xi_{k_1} \xi_{k_2} \xi_{k_3} \dots,  \qquad
\text{where $k_j=-j$ for sufficiently large $j$}
.$$
We allow to apply the rule (4.1) finite number
of times.
Hence, for $\sigma\in S_{2\infty}$,
$$
\Xi_{\sigma \frk}=(-1)^\sigma \Xi_\frk
.$$
Modulo permutations,  each collection $\frk$ can be reduced to the form
\begin{equation}
\frk:\, k_1>k_2>k_3>\dots, \qquad
\text{where $k_j=-j$ for sufficiently large $j$}
.\end{equation}

We emphasis that
$$
\Bigl\{
\text{number of $k_j\ge 0$}
\Bigr\} =
\Bigl\{
\text{number of negative $l$ such that $l\ne k_i$ $\forall i$}
\Bigr\}
.$$

We define the space $\L$ as the Hilbert space whose
orthonormal basis is $\Xi_\frk$, where $\frk$ satisfies (4.2).
We also define the space $\ov\L$ whose elements are formal series
$\sum_\frk c_\frk\Xi_\frk$ with arbitrary $c_\frk$.


\smallskip

{\bf \punkt Remark: the usual fermion Fock space.}
 A fermion Fock space is a  space of functions depending
on a countable collection of anticommuting variables.

Consider an infinite
   collection
$\xi_0$, $\xi_1$, \dots, $\eta_1$, $\eta_2$,\dots
of anticommuting variables.
Consider  all possible finite monomials
\begin{equation}
\xi_{u_1}\dots \xi_{u_n}\, \eta_{v_1}\dots \eta_{v_m};
\qquad
\end{equation}

Denote by $\cL$  the space ({\it the fermion Fock space}),
 whose orthonormal basis
consists of such monomials.

 To each  vector $\Xi_\frk$, we assign
the vector
$$ \prod_{k_j\ge 0} \xi_{k_j}
\prod_{l:\, l<0, l\ne k_i\,\, \forall i} \eta_{-l}
.$$

Hence we obtain a space, whose basis consists of products
$\xi_{u_1}\dots \xi_{u_n}\, \eta_{v_1}\dots \eta_{v_n}$.
As we observed above, the number of $\xi$ in this product
 equals the number of $\eta$; i.e., we have $m=n$ in (4.3)%
\footnote{In particular, to each sequence $\scriptstyle\frk$ we associated
a collection of integers $u_1$,\dots, $u_n$, $v_1$, \dots, $v_n$.
This is the Frobenius parameters of Young diagrams, see \cite{Mac},I.1.}
.

Thus, $\L$ is a subspace in $\cL$.

\smallskip

{\bf \punkt Decomposable vectors
 in the Hilbert space $\L$.}
Consider two infinite  matrices $A=\{a_{mi}\}$,
$B=\{b_{mj}\}$, where
$m >0$, $0 \le i <\infty$, $-\infty<j <0$.
Assume that $A$ is a Hilbert--Schmidt matrix and
$B$   belongs to the trace class%
\footnote{This condition can be replaced by other variants,
see a discussion below and the next subsection.}.

Consider the product
$$
\Xi[A;1+B]:=
\prod_{m=1}^{\infty}
\bigl(\xi_{-m}+
\sum_{i\ge 0}   a_{mi} \xi_{i}+\sum_{j<0} b_{mj}\xi_j  \bigr)
;$$
%
removing parentheses, we choose the summand $\xi_{-m}$
from each parenthesis except a finite number of factors.

We call $\Xi[A,1+B]$ by {\it decomposable vectors}.
The vectors $\Xi_\frk$ defined above also have the form $\Xi[A,1+B]$.

\smallskip

{\sc Remark.} Equivalently, we can consider one matrix
$R:=\begin{pmatrix}A & 1+B\end{pmatrix}$, whose matrix elements
$r_{ml}$ are indexed by $m\ge 1$, $l\in \Z$. After this, we can write
$$\Xi[R]:=\prod_{m=1}^\infty \bigl(\sum_l r_{ml}\xi_l\bigr)$$
 We prefer
$\Xi[A;1+B]$ as a basic notation, since it is
more convenient for understanding of convergence/divergence
of our series.

\smallskip

{\sc Proposition \utver}
a) $ \Xi[A,1+B]\in \L$.

\smallskip

b) $ \langle \Xi[A_1,1+B_1], \Xi[A_2,1+B_2]\rangle_\L=
   \det\bigl(A_1 A_2^* +(1+B_1)(1+B_2^*)\bigr)$.

\smallskip

c) {\it $ \Xi[A, B]=\sum_\frk \gamma_\frk \Xi_\frk$, where
 $\gamma_\frk$ is the minor of the matrix
$\begin{pmatrix} A& 1+B\end{pmatrix}$ consisting of
 the columns with numbers $k_1$, $k_2$, \dots.}

\smallskip

d) {\it Let $D$ be a matrix of the trace class. Then}
$$\Xi[(1+D)A, (1+D)(1+B)]=\det(1+D)\, \Xi[A, B].$$

\smallskip

{\sc Proof.}
This statement is a variant of Lemma 4.1.
The minors in Statement b) converge since $B$ is in the trace class.
   Thus we obtain a well-defined
formal series in vectors $\Xi_\kk$.
We have $\|\Xi[A;1+B]\|^2=\det(AA^*+(1+B)(1+B)^*)$.
This determinant is convergent since
$AA^*$ and $B$ are  in the trace class.
\hfill $\square$

\smallskip

Thus different matrices
$\begin{pmatrix} A& 1+B\end{pmatrix}$
can give the same vector $\Xi[A,1+B]$.
To reduce this freedom, we can consider the system
of vectors $\Xi[D,1]$
\begin{equation}
\prod_{m=1}^\infty\bigl(\xi_{-m}+\sum_{i=0}^\infty d_{mi}\xi_i\bigr)
\end{equation}
 A representation of
a vector $\Xi[A, 1+B]$ in the form $\mathrm{const}\cdot\Xi[D,1]$ is unique
(indeed $D=(1+B)^{-1}A$), but in the case
$\det(1+B)= 0$ such representation does not exist
(it does not exist for all the vectors $\Xi_\frk$
except $\Xi_{-1,-2,\dots}$).

In the notation of 4.3,  element (4.4) corresponds to
$$
\exp\bigl\{\sum\nolimits_{mi} d_{mi}\xi_i \eta_m\bigr\}
,$$
 it is also a spinor function in the terminology
of \cite{Ner-categories}.

\smallskip

{\bf\punkt Decomposable vectors in the space of
formal series $\ov\Lambda$.}
 Let $A$ be an arbitrary matrix
and $B$  satisfies the condition
\begin{equation}
\text{$b_{mj}=0$ for $j\le -m$}
\end{equation}
 Then the vector
$\Xi[A; 1+B]$ is a well-defined formal series in
vectors $\Xi_\frk$, i.e., $\Xi[A; 1+B]\in\ov\L$.
In other words,
expressions of the following type are well defined
\begin{equation}
\prod_{m=1}^\infty\bigl(
 \xi_{-m}+c_{m1} \xi_{-m+1} +c_{m2} \xi_{-m+2}+\dots\bigr)
\end{equation}
for arbitrary $c_{mj}$.

Indeed, all the coefficients
of its expansion in the basis $\Xi_\frk$
(see Proposition 4.2.c)
are determinants having the following structure:
diagonal elements are 1 starting some place, and
only finite number of entries upper the diagonal
are nonzero. Hence an evaluation of our coefficients
is reduced to an evaluation of some
 determinants of finite size.

\smallskip

The expansion of (4.6) contains the term $\xi_{-1}\xi_{-2}\dots $
and hence there are vectors $\Xi[A; 1+B]$ that can not be represented
in this form.

\smallskip

 More generally, let $A$ be arbitrary
and $B$ satisfy the condition
\begin{equation}
\text{$b_{mj}=0$ for $j\le -m$ except finite number of entries}
\end{equation}
In other words, we consider products of the form
$$
\prod_{m=1}^\infty\bigl(
 \xi_{-m}+c_{m1} \xi_{-m+1} +c_{m2} \xi_{-m+2}+\dots+
\sum_j\sigma_{mj}\xi_j\bigr)
,$$
where $\sigma_{mj}=0$ except finite number of entries.
Again, we obtain a well-defined element
of the space $\ov\L$.

  Below the both variants of the definition
of decomposable vectors (in $\Lambda$ and in $\ov\Lambda$)
are sufficient for our purposes.

\smallskip

{\bf \punkt The action of $\GL_\infty$ in $\L$.}
We consider infinite matrices
$H=\{H_{ij}\}$, where $i$, $j$ range in $\Z$.
Let $\delta_{ij}$ be the Kronecker symbol, i.e.,
$\delta_{ii}=1$ and  $\delta_{ij}=0$ for $i\ne j$.

We define the group   $\GL_\infty$ as the group
of all invertible matrices $H$ such that
$h_{ij}-\delta_{ij}=0$ for all pairs $(i,j)$
except finite number of entries.

We define the representation $\rho$ of $\GL_\infty$ in $\L$
by the assumption
$$\xi_i\mapsto \sum_i h_{ij} \xi_j.$$
The vectors $\Xi_\frk$
and their linear combinations
 are transformed corresponding this rule.

We have
\begin{equation}
\rho(H)\Xi_\frk=
\sum_{\frl:\,\, l_1>l_2>\dots;\,\,\, \text{$l_j=-j$ for large $j$}}
\det\limits_{i,j\ge 1}
\{h_{k_i l_j}\}\,\, \Xi_\frl
.\end{equation}
This equality can be considered as a definition of the operators
$\rho(H)$.

The operators $\rho(H)$ transform decomposable vectors
 to decomposable vectors:
$$\rho(H) \Xi[ A, 1+B]=\Xi[C, 1+D],\qquad
\text{where
$\begin{pmatrix} C&1+ D\end{pmatrix}=
  \begin{pmatrix} A& 1+B\end{pmatrix} H$}
.$$

{\bf \punkt The  space of skew-symmetric functions.}
Let $x_1$, $x_2$, \dots be formal variables.
We say that a {\it long monomial} is an infinite product
$$
x_1^{k_1}x_2^{k_2}x_3^{k_3}\dots,
 \qquad\text{where $k_j=-j$ for large $j$}
.$$

We say that a formal linear
combination of long monomials
is a  {\it skew-symmetric function}
if it is
 skew-symmetric with respect
to finite permutations of variables.

Fix a collection
$$\frk: \quad k_1>k_2>k_3>\dots
\qquad\text{where $k_j=-j$ for large $j$}
.$$
Define the skew-symmetric function
$$
\Omega_\frk=\sum_{\sigma\in S_\infty}
(-1)^\sigma
  x_{\sigma(1)}^{k_1}x_{\sigma(2)}^{k_2}x_{\sigma(3)}^{k_3}\dots
$$
We introduce the scalar product in the space of skew-symmetric
functions by the assumptions:

1. $\Omega_\frk$ are pairwise orthogonal

2. $\|\Omega_\frk\|=1$.

Denote by $\A$
the Hilbert space of skew symmetric functions, i.e.,
the space, whose elements are series
$$
\sum\nolimits_\frk c_\frk \Omega_\frk,
\qquad \text{where $\sum\nolimits_\frk |c_\frk|^2<\infty$}
.$$

We also define the space $\ov \A$, whose elements are
formal series
$\sum_\frk c_\frk \Omega_\frk$ without any conditions
for the coefficients $c_\frk$.

{\bf \punkt Variant of definition.}
Consider a countable collection of
formal variables $x_1$, $x_2$, \dots.
We consider formal series (in the usual sence)
in $x_j^{\pm 1}$ satisfying the following condition
of skew symmetry
$$
f(x_1,x_2,\dots, x_i,\dots, x_j, \dots)=
-\left(\frac{x_j}{x_i}\right)^{j-i}
f(x_1,x_2,\dots, x_j,\dots, x_i, \dots)
.$$
Then
$f(x)\cdot \prod_{j=1}^\infty x_j^{-j}$
is a skew symmetric function.

\smallskip

{\bf \punkt Identification of $\L$ and $\A$.}
The canonical unitary operator $\L\to\A$
is defined by
$\Xi_\frk \mapsto\Omega_\frk$.

\smallskip

{\bf \punkt Images of decomposable vectors.}
Fix matrices $A$, $B$ as in 4.4 or  in 4.5.
For each $m=1,2,\dots$ we consider the formal Laurent
series
$$
q_m(x)=\sum\nolimits_{i\ge 0} a_{mi} x^i +\sum\nolimits_{j<0} b_{mj} x^j
.$$
Also we will use an alternative notation for the same series
$$
q_m(x)=\sum\nolimits_{-\infty<p<\infty} \kappa_{mp}x^p
$$

Consider the determinant
$$
\Omega[A,1+B]=
\det\begin{pmatrix}
 x_1^{-1} +q_1(x_1)&x_2^{-1} + q_1(x_2) &x_3^{-1} + q_1(x_3) &\dots\\
 x_1^{-2}+  q_2(x_1)& x_2^{-2}+ q_2(x_2) &x_3^{-2}+ q_2(x_3) &\dots  \\
 x_1^{-3}+      q_3(x_1)&x_2^{-3}+  q_3(x_2) &x_3^{-3}+ q_3(x_3) &\dots   \\
                   \vdots &  \vdots & \vdots &           \ddots
   \end{pmatrix}
.$$
We understand this determinant as the following  series
in long monomials
\begin{equation}
\sum_{n=0}^\infty
\sum_{1\le k_1<k_2<\dots<k_n}
              \Bigl\{
\det\limits_{1\le\alpha,\beta\le n,\, }
\{x_{k_\beta}^{-k_\alpha} (1-\delta_{\alpha\beta})+ q_{k_\alpha}(x_{k_\beta})\}
\prod_{j:j\ge 1, j\ne k_\alpha} x_j^{-j}
\Bigr\}
.
\end{equation}
This means that we consider only summands of the determinant that
differ from the diagonal product $x_1^{-1} x_2^{-2} x_3^{-3}\dots$
in finite number of terms.
Under our conditions for $A$, $B$, this series converges.

Another variant to understand this determinant is
\begin{equation}
\lim_{N\to\infty}\Bigl[ \det_{1\le j,k\le N}
   \{x_k^{-m}+q_m(x_k)\}\cdot \prod_{l=N+1}^\infty x_l^{-l}
\Bigr]                    ;
\end{equation}
the limit means the coefficient-wise limit%
\footnote{The expression in brackets
 is a series in long monimials
but not a skew symmetric function.}.

If $B=0$ and $A$ is arbitrary,
 then our series is a well defined
formal series, i.e., it is an element of $\ov \A$.
The same is valid if $B$ satisfies condition (4.5) or
condition (4.7).  For this case, the series
for any coefficient in (4.9) is finite
and each sequence of coefficients in (4.10)
is stabilized
starting some place.

In particular,
the determinants of the form
$$
\det\{x_k^{-m}(1+\sum_{j>0} c_{mj} x_k^j)\}
$$
are well-defined  elements of $\ov \A$.

\smallskip

{\sc Proposition \utver} {\it  The canonical map $\L\to\A$
 takes a vector $\Xi[A,1+B]\in\L$
to the vector $\Omega[A,1+B]\in \A$.}

\smallskip

{\sc Proof.} It is sufficient to expand $\Omega[A; 1+B]$
in the functions $\Xi_\kk$.
The coefficient at $\Xi_\kk$  is
$$
\det\limits_{1\le m\le\infty, 1\le\alpha\le\infty}
\{\kappa_{mk_\alpha}+\delta_{-m,k_\alpha}\}
$$
It coincides with the corresponding coefficient
 in Proposition 4.2c.

{\sc Remark.}
Also, we can write the following expression for $\Omega[A; 1+B]$
$$
x_1^{-1} x_2^{-2} x_3^{-3}\dots
\det\begin{pmatrix}
  1 +x_1q_1(x_1)      &x_2 + x_2^{2}\,q_1(x_2)        &x_3^{2} +x_3^{3} \, q_1(x_3) &\dots\\
 x_1^{-1}+ x_1 q_2(x_1)& 1+x_2^{2}\, q_2(x_2)         &x_3+x_3^{3}\,  q_2(x_3) &\dots  \\
 x_1^{-2}+ x_1 q_3(x_1)& x_2^{-1}+ x_2^{2}\, q_3(x_2) & 1+x_3^{3}\,  q_3(x_3) &\dots   \\
                   \vdots &  \vdots & \vdots &           \ddots
   \end{pmatrix}
$$
and consider the determinant as a determinant $\det(1+Z)$
in the sence of formal series in $x_j^{\pm1}$.
This is equivalent to (4.9).

\medskip

{\bf \large 5. Fermion-symmetric correspondence}

\nopagebreak

\medskip

\nopagebreak

\addtocounter{sec}{1}

\setcounter{equation}{0}
\setcounter{punkt}{1}
\setcounter{utver}{1}

Here we consider only the
Redfield scalar product in $\S$, i.e.,
we assume $\omega_j=j$.
We denote our space  $\S_\omega$
 by $\S_{\cl}$.

\smallskip

{\bf \punkt The canonical unitary operator $\S_\cl\to\A$.}
Consider the element
$\Delta\in\A$ defined  by
$$\Delta:=\Omega_{-1,-2,-3,\dots}=
\sum_{\sigma\in S_\infty} (-1)^\sigma
x_{\sigma(1)}^{-1}  x_{\sigma(2)}^{-2}   x_{\sigma(3)}^{-3} \dots
$$

For $f\in \S$, we consider the skew-symmetric
function $\Delta\cdot f$.

\smallskip

{\sc Proposition \utver} {\it For the Schur functions $s_\frk$,}
$$\Delta s_\frk= \Omega_\frk.$$

{\sc Proof.} For  fixed  $N$, let us find terms
(long monomials)  of $\Delta\cdot s_\frk$
that contain the factor $\prod_{j>N} x_j^{-j}$.
For this, we must follow terms of $s_\frk$, that do not contain
$x_{N+1}$, $x_{N+2}$, \dots.
Hence it is sufficient to evaluate
$$
s_\frk \Bigl|_{x_{N+1}=x_{N+2}=\dots=0}\cdot
 \sum_{\sigma\in S_n}(-1)^\sigma
\prod_{j=1}^N x_{\sigma(j)}^{-j}.
$$
But,
$$
\sum_{\sigma\in S_N}(-1)^\sigma\prod_{j=1}^N x_{\sigma(j)}^{-j}
 =
\prod_{1\le i<j\le N} (x_i-x_j)
\prod_{j=1}^N x_j^{-N}
$$
and this  implies the required statement. \hfill $\square$.

\smallskip

{\sc Corollary \utver} a) {\it The map $f\mapsto \Delta f$
is a unitary operator $\A\to\S_\cl$.}

\smallskip

b)
{\it The map $f\mapsto \Delta f$   is a well defined map
of spaces of formal series $\ov\S\to\ov\A$.}

\smallskip

{\bf \punkt Preimage of decomposable vectors.}

\nopagebreak

\smallskip

{\sc Lemma \utver} (\cite{Hua}).
{\it Let $r_m(z)=c_0^{(m)} + c_1^{(m)} z +  c_2^{(m)} z^2+\dots$
be formal series, $m=1,\dots,N$. Then}
$$
\det_{1\le m\le N, 1\le p\le N}\{r_m(z_p)\}=\sum_{l_1<\dots< l_N}
\det_{1\le m\le N, 1\le j\le N}
    \{c^{(m)}_{l_j}\}
  \det_{1\le j\le N, 1\le p\le N}
     \{z_p^{l_j}\}
.$$

Let $A$ be arbitrary and $B$ satisfies the condition (4.7).
Fix $M$ such that $b_{mi}=0$ if $i\ge m$, $i\ge M$.

 Consider the symmetric function
$\Pi[A,1+B]$ defined by the condition
\begin{multline*}
\Pi[A,1+B]\Bigr|_{x_{N+1}=x_{N+2}=\dots=0}
= \\ =
\frac{\prod_{p=1}^N x_p^N\cdot \det_{1\le m,p\le N}
\{x_p^{-m}+\sum_{i\ge 0} a_{mi} x_p^i +
\sum_{1\ge j \ge -N} b_{mj} x_p^j\}}
{\prod_{1\le p<q\le N} (x_p-x_q)}
\end{multline*}
for all $N>M$.

\smallskip

{\sc Corollary \utver}
$$\Delta\cdot \Pi[A,1+B]=\Omega[A,1+B].$$

{\sc Proof.} We expand $\Pi[A,1+B]$ in the Schur functions
using Lemma 5.3 and expand $\Omega[A,1+B]$ in functions
$\Xi_\kk$ using Proposition 4.2c. We observe that the coefficients
of these two expansions coincide. It remains to apply
Proposition 5.1.

\smallskip

{\bf \punkt The action of $\GL_\infty$ in $\S_\cl$.}

\nopagebreak

\smallskip

{\sc Lemma \utver}
{\it Let $R(x,y)=\sum_{ij} c_{ij} x^i y^j$ be a formal series.
Then}
$$
\det_{\substack{1\le k \le N,\\ 1\le l \le N}} \{R(x_k,y_l)\}
=
\sum_{\substack{i_1<\dots <i_N,\\ j_1<\dots <j_N}}
\det_{\substack{1\le \alpha \le N, \\ 1\le \beta \le N }}
   \{c_{i_\alpha j_\beta}\}
\det_{\substack{1\le k\le N\\ 1\le \alpha\le N} }
   \{ x_k^{i_\alpha}\}
\det_{\substack{1\le l\le N\\ 1\le \beta\le N }}
   \{ y_l^{j_\beta}\}
.$$

{\sc Proof.} First, our expression
is skew symmetric with respect to $x_k$
and also skew symmetric with respect to
$y_l$. Hence it has the form
$$
\sum \gamma(i_1,\dots,i_N;j_1,\dots, j_N)
\det  \{ x_k^{i_\alpha}\}
\det   \{ y_l^{j_\beta}\}
.$$
A coefficient $\gamma$ coincides with the coefficient
at $\prod x_\alpha^{i_\alpha} \prod      y_\beta^{j_\beta}$
in $\det\{R(x_k,y_l)\}$.
The latter coefficient can be easily evaluated.
\hfill $\square$.

\smallskip


Let $H=\{h_{ij}\}$ be an element
of $\GL_\infty$, see 4.6. Let an integer  $M$
be sufficiently large such that for
$i<-M$ we have $h_{ij}-\delta_{ij}=0$
and the same holds for $j<-M$.
We define the bisymmetric kernel $L[H]$
by
\begin{multline*}
L[H](x,y)\Bigr|_{\substack{x_{N+1}=x_{N+2}= \dots=0\\
                          y_{N+1}=y_{N+2}= \dots=0 }}
=\\=
\frac{
    \det_{1\le k\le N, 1\le l\le N}
    \{ \sum_{i\ge -N,j\ge -N} h_{ij}x_k^i y_l^j\}
    \prod_{j=1}^N x_j^N y_j^N
              }
   {\prod_{1\le p<q\le N} (x_p-x_q)
   \prod_{1\le p<q\le N} (y_p-y_q) }
\end{multline*}
for all $N>M$.
Let $\cA[H]$ be the operator
 defined by the bisymmetric kernel
$K[H]$.

Lemma 5.5 implies the following statement.

\smallskip

{\sc Corollary \utver} {\it The following expansion in the Schur
functions holds
$$
K[H](x,y)
=
\sum_{\fri, \frj}
\det_{1\le \alpha <\infty,  1\le \beta <\infty }
  \{h_{i_\alpha j_\beta}\}
 s_\fri(x) s_\frj(y)
,$$
 where the summation is taken over all the sequences
$\fri$: $i_1>i_2>\dots$,
and $\frj$:  $j_1>j_2>\dots$, where $i_\alpha=-\alpha$
for sufficiently large $\alpha$ and
$j_\beta=-\beta$ for sufficiently large $\beta$.}

\smallskip

{\sc Remark.}  All the minors $\{h_{i_\alpha j_\beta}\}$
satisfy the property:
$h_{i_\alpha j_\beta}-\delta_{\alpha\beta}=0$
for all pairs $(\alpha,\beta)$ except finite number.

{\sc Theorem \utver}
a) {\it The map $H\mapsto \cA[H]$
is a representation of the group $\GL_\infty$, i.e.,
 $\cA[H]\cA[G]=\cA[HG]$.}

\smallskip

b) {\it The canonical unitary operator $\L\to\A\to\S_\cl$
intertwines the representation $\rho(H)$ given by (4.8)
and the representation $\cA[H]$.}

\smallskip

{\sc Proof.} It is sufficient to verify b).
We must check that the matrix elements of $\rho[H]$
in the basis $\Xi_\frk\in\L$ are equal to
the matrix elements of $\cA[H]$
in the basis $s_\frk\in\S_\cl$. The  matrix elements
of $\rho[H]$ are
given by (4.8). To find the  matrix elements of $\cA[H]$,
we must  expand the kernel $L[H]$ in
the the Schur functions,
see Proposition 1.5. This expansion
is given by Corollary 5.6.
\hfill $\square$

\smallskip

{\sc Remark.} Let extend the formalism of kernels to the space
$\A$. Consider a series
$$
\sum_{k_1,k_2,\dots;l_1,l_2,\dots} a_{k_1,k_2,\dots;l_1,l_2,\dots}
x_1^{k_1} x_2^{k_2}\dots y_1^{l_1} y_2^{l_2}\dots
$$
in long monomials. We say that it is a {\it biskewsymmetric kernel}
if it is skew symmetric with respect to $x_j$ and with respect to
$y_j$. For a biskewsymmetric kernel $K(x,y)$,
we define the operator $A:\Lambda\to\Lambda$ by
$$
Af(x)=\langle K(x,y),f(y)\rangle_{\Lambda[y]}
$$
The group $\GL_\infty$ acts in $\A$ by operators whose
biskewsymmetric kernels are
$$
K(x,y)=\det_{kl}\{\sum_{i,j} h_{ij}x_k^i y_l^j\}
,$$
we consider only terms of the determinant that differ
from the product $\prod\nolimits_{k}^\infty (x_k y_k)^{-k}$
in a finite number of places, see 4.10.

\smallskip

{\bf \punkt Images of some multiplicative vectors.}
  Let $r(x)= 1+ r_1 x +r_2 x^2+\dots$.
Consider the vector
\begin{equation}
R(x)=\prod\nolimits_{k=1}^\infty r(x_k)\in\ov\S.
\end{equation}

{\sc Proposition \utver}
 a) {\it The image of the vector $R$
under the maps $\S_\cl\to \A\to \Lambda$ is
\begin{equation}
\prod_{m=1}^\infty
 \bigl(\xi_{-m}+r_1 \xi_{-m+1}+r_2 \xi_{-m+2}+\dots\bigr)
.\end{equation}
If $R(x)\in\S_\cl$, then (5.2) is an element of $\Lambda$.
Otherwise, (5.2) is in $\ov\Lambda$.}

\smallskip

 b) {\it The image of the vector $R$ under the map
$\S_\cl\to \A$ is given by
$$
\det\{x_k^{-m} r(x_k)\}
,$$
where the determinant is defined in the same way as
in 4.10.}

\smallskip

{\sc Proof.}
The statement a) is a corollary of b).
To obtain b),  we write
$$
\prod_{k=1}^N r(x_k)=
   \frac{\{\det x_k^{m-1} r(x_k)\}}
    {\prod_{1\le l<k\le N}(x_k-x_l)}
$$
and tend $N$ to infinity.
\hfill $\square$

\smallskip

{\bf \punkt Another expressions for the same vectors.}
Let $r(x)$ and $R$ be the same as above (5.1). We intend to
write the canonical form (4.4)
for the image of $R$ in $\Lambda$.
Let
$$
r(x)=1+r_1 x+r_2x^2+\dots =\exp a(x) =\exp\{a_1 x+a_2 x^2+\dots\}
$$
By Proposition 3.1a, we have $R\in\S_\cl$ iff $\sum j|a_j|^2<\infty$.
Therefore, $a(x)$ is holomorphic in the disc $|x|<1$,
and hence $r(x)$ is holomorphic in the same disc.
Consider the function
$$ \frac{r(x)/r(u)-1} {x-u}.$$
This function is holomorphic in the bidisc $|x|<1$, $|u|<1$,
and hence it can be expanded in a power series
$$
 \frac{r(x)/r(u)-1} {x-u}=
\sum_{\alpha\ge 0, \beta\ge 0}
    \zeta_{\alpha\beta} x^\alpha u^\beta
.$$

{\sc Proposition \utver}
a) {\it The image of $R$ under the map $\S_\cl\to\A\to\L$
is }
$$
\prod_{m=1}^\infty
\bigl(\xi_{-m}+\sum_{j\ge 0} \zeta_{(m-1)j} \xi_j\bigr)
$$

b) {\it The image of $R$ under the map $\S_\cl\to \A$ is}
$$
\det_{m \ge 1,\, k\ge 1}
\bigl\{ x_k^{-m}+\sum_{j\ge 0} \zeta_{(m-1)j} x_k^j\bigr\}
$$

{\sc Proof.} It is sufficient to prove b).
By Proposition 5.8, the required  vector is
the determinant of the matrix
$$
\begin{pmatrix}
\dots+r_2 x_1+r_1 + x_1^{-1}&
\dots+r_2 x_2+r_1 + x_2^{-1}&
\dots+r_2 x_3+r_1 + x_3^{-1}&
\dots \\
\dots+r_2 +r_1x_1^{-1} + x_1^{-2}&
\dots+r_2 +r_1x_2^{-1} + x_2^{-2}&
\dots+r_2 +r_1x_3^{-1} + x_3^{-2}&
\dots \\
\dots+r_2 x_1^{-1} +r_1x_1^{-2} + x_1^{-3}&
\dots+r_2 x_2^{-1} +r_1x_2^{-2} + x_2^{-3}&
\dots+r_2 x_3^{-1} +r_1 x_3^{-2} + x_3^{-3}&
\dots \\
\vdots&\vdots&\vdots&\ddots
\end{pmatrix}
$$
We can replace $n$-th row by a linear
combination of the first $(n-1)$ rows.
Hence, we can replace the series in the $n$-th row by
$$
\mu_n(x)=x^{-n}(1+r_1x+r_2 x^2+\dots)
  (1+s_1 x+ s_2 x^2+ \dots +s_{n-1}x^{n-1})
,$$
where $s_j$ are arbitrary.

Let us choose $s_j$ defined by the rule
$$
(1+r_1x+r_2 x^2+\dots) (1+s_1 x+ s_2 x^2+ \dots )=1
$$
Then
$$
\mu_n(x)=x^{-n}+\sum_{j\ge 0} C_{nj} x^j
$$
where
$$C_{nj}=r_{n+j}+ r_{n+j-1}s_1+\dots +r_{j+1}v_{n-1} $$
We write the generating function  $Q(x,y)=\sum C_{nj}x^{n-1}y^j$.
A direct calculation gives
$Q(x,y)(x-y)=r(x)/r(y)-1$.
\hfill $\square$

\smallskip

{\bf \punkt  Example:
inversion formula for boson-fermion correspondence.}
A {\it boson-fermion correspondence}%
\footnote{In this definition, the boson Fock space $\F$
is identified with the subspace $\Lambda$ in the fermion
Fock space $\cL$. The whole space $\cL$ can be identified with
$\F\otimes \C[t,t^{-1}]$, where $\C[t,t^{-1}]$ is space
of Laurent series in formal variable $t$, and
$\langle t^k, t^l\rangle=\delta_{kl}$. There are
some other variants of this correspondence, see
\cite{Fre}, \cite{PS}, \cite{MJD}, \cite{Ner-categories}
for further discussions; discovering of the correspondence
usually is attributed to Skyrme, 1971.}
  is the composition
of the canonical maps
$$
\F\to\S_\cl\to\A\to\Lambda
.$$
Obviously, it is a unitary operator.

Let $f(z)$ be an element of $\F$, let $g(\xi)$
be the corresponding element of $\Lambda$.
We intend to write a formula that reconstructs
$f$ by $g$. For this, we find the images in $\Lambda$ of
the vectors
$\phi_a\in\F$, see (2.4), and apply the reproducing
property (2.5), see proof of Proposition 3.1.

Define the polynomials $\cR_n(z)$ by
$$
\exp \bigl\{z_1x+z_2x^2+z_3x^3+\dots \bigr\}
=1+\cR_1(z)x+\cR_2(z)x^2+\dots
.$$
In other words,
\begin{equation}
\cR_n(z)=
\sum_{s_1+2s_2 +3s_3+\dots =n}
\prod_j \frac {z_j^{s_j}} {s_j!}
.\end{equation}
It will be also convenient to assume
\begin{equation}
\cR_0=1,\,\, \cR_{-1}=\cR_{-2}=\dots=0
.\end{equation}

We evaluate the image of $\phi_a$ in $\Lambda$
 by Proposition 5.8 and obtain
$$
f(\sqrt 1 z_1,\sqrt 2 z_2,\sqrt 3 z_2, \dots))=
\bigl\langle g(\xi),\,
 \prod_{m=1}^\infty
\bigl(\xi_{-m}+\cR_1(\ov z)\xi_{-m+1}+ \cR_2(\ov z)\xi_{-m+2}+
   \dots\bigr)\bigr\rangle_\Lambda
$$


{\bf \punkt  Example:
another inversion formula for the boson-fermion correspondence.}
Proposition 5.9 allows to obtain another formula.
Define the polynomials
$\cQ_{mn}(z)$  by
\begin{equation}
\frac{ \exp\bigl\{\sum_{j>0} z_j(x^j-y^j)\bigr\}-1}
     {x-y}=\sum \cQ_{mn}(z)x^m y^n
\end{equation}
Then
$$
f(\sqrt 1 z_1,\sqrt 2 z_2,\sqrt 3 z_2, \dots)=
\bigl\langle g(\xi),\,
 \prod_{m=1}^\infty
\bigl(\xi_{-m}+ \sum_{j=0}^\infty \cQ_{(m-1)j}(z) \xi_j
   \bigr)\bigr\rangle_\Lambda
.$$

{\bf \punkt Example: generating functions for
 characters
of the symmetric groups.}
Let $\mu:\,\mu_1\ge\mu_2\ge\dots>\mu_h>0$ and
$\lambda: \lambda_1\ge\lambda_2\ge\dots> 0$;
let $\sum\mu_j=\sum \lambda_j=N$.
Consider the irreducible
 representation of the symmetric group $S_N$ corresponding
the Young diagram $\mu$;
denote by $\chi^\mu_\lambda$ the value of its
character on a permutation whose cycles has
lengths $\lambda_j$. Then (see \cite{Mac}, I.7.7)%
\footnote{In particular, this gives images of the
functions $e_\ll(z)\in\F$ in $\Lambda$ and
preimages of the vectors $\Xi_\frm\in\Lambda$
in $\F$. This was widely used in matematical physics.}
$$
\chi^\mu_\lambda=
     \langle p_\lambda, s_\mu\rangle
$$
(in notation 1.2).
Let $l_j$ be the number of entries of $j$ into
the collection $\lambda$,  let
$m_j=\mu_j-j$, and let $\frm:=(m_1,m_2,\dots)$. We denote
$\chi^\frm_\ll:=\chi^\mu_\lambda$
and write
$$
\chi^\frm_\ll = \langle p_\ll, s_\frm  \rangle;
\qquad  p_\ll=\sum_\frm \chi^\frm_\ll s_\frm
$$
Hence
\begin{equation}
\exp\bigl\{ \sum_{j>0} a_j p_j(x)\}=
\sum_\ll
\Bigl(\prod_j \frac{a_j^{l_j}} {l_j!}\Bigr)p_\ll
=
\sum_\ll \sum_\frm
\Bigl(\prod_j \frac{a_j^{l_j}} {l_j!}\Bigr)\chi^\frm_\ll \cdot s_\frm
\end{equation}

Then we transform the left-hand side to the form
$$
\prod_k\exp\bigl\{\sum a_jx_k^j\bigr\}=
\prod_k\bigl\{\sum_{j>0} \cR_j(a)x^j_k  \bigr\}
$$
Multiplying the left hand side and the right hand side of (5.6)
 by $\Delta$, we obtain
\begin{equation}
\det\{x_k^{-m} (1+\sum_{j>0} \cR_j(a) x^j_k)\}
=
\sum_\ll \sum_\frm\Bigl( \prod_j \frac{a_j^{l_j}} {l_j!}\Bigr) \cdot
\chi^\frm_\ll \Omega_\frm
\end{equation}
Now we are ready to write 3 formulae for generating functions.

1. Expanding the left hand side of the last equality
in $\Omega_\frm$, and equating its coefficients,
 we obtain the following generating function
for values of the character of a given representation
of a symmetric group $S_N$
           $$
\sum_\ll\Bigl( \chi_\frm^\ll \prod_j \frac{a_j^{l_j}} {l_j!} \Bigr)
  =
\det_{1\le i,j\le h} \{ \cR_{m_i-j}(a)\}
             ,$$
the polynomials $\cR$ are defined by (5.3)--(5.4) and
 $m_{k}=-k$ for $k>h$.

2. Applying the construction of 5.7, we obtain another
 expression  for the same generating function.
   Let $u_1$,\dots,$u_n$,
$v_1$,\dots,$v_n$ are the parameters for $\frm$
(or $\mu$) defined in 4.3.    Then
           $$
\sum_\ll\Bigl( \chi_\frm^\ll \prod_j \frac{a_j^{l_j}} {l_j!} \Bigr)
  =
\det_{1\le i,j\le n}\{\cQ_{v_{j-1} u_i}(a)\}
$$
where the polynomials $\cQ$ are defined by (5.5).

3.
We also write identity (5.7) in the form
$$
\Bigl(\sum_{\sigma\in S_\infty} (-1)^\sigma
\prod_{k=1}^\infty x_{\sigma(k)}^{-k}\Bigr)
\cdot
\prod_{k=1}^\infty \exp\Bigl\{\sum_{j=1}^\infty a_j x_k^j\Bigr\}
=
\sum_\ll \sum_\frm\Bigl\{
 \chi_\frm^\ll
\cdot
          \Bigl( \sum_{\sigma\in S_\infty} (-1)^\sigma
        \prod_{k=1}^\infty x_{\sigma(k)}^{m_k}\Bigr)
\cdot
 \prod_j \frac{a_j^{l_j}} {l_j!}
\Bigr\}
$$
Thus, the coefficients of
 the formal series in the left hand side
are all the values of all the characters of all
the symmetric groups
(up to    signs and 
 products of factorials).

\sc
Math.Phys Group

Institute for Theoretical and Experimental Physics

B.Cheremushkinskaya

Moscow 117 259 Russia

\tt neretin@mccme.ru

    neretin@gate.itep.ru

\end{document}